\documentclass[twocolumn]{aastex63}

\accepted{\today}
\submitjournal{ApJ}

\usepackage{comment}
\usepackage{makecell}
\usepackage{amsmath}
\usepackage{longtable}

\begin{document}

\author[0000-0002-8770-6764]{R\'eka K\"onyves-T\'oth}
\affiliation{Konkoly Observatory, Research Centre for Astronomy and Earth Sciences, H-1121 Budapest Konkoly
Th. M. út 15-17., Hungary; MTA Centre of Excellence}
\affiliation{Department of Optics \& Quantum Electronics, University of Szeged, D\'om t\'er 9, Szeged, 6720 Hungary}
\affiliation{ELTE Eötvös Loránd University, Gothard Astrophysical Observatory, Szombathely, Hungary}

\shorttitle{Pre-maximum SLSNe-I}
\shortauthors{K\"onyves-T\'oth, R}

\correspondingauthor{R\'eka K\"onyves-T\'oth}
\email{konyvestoth.reka@csfk.mta.hu}

\graphicspath{{./}{}}

\title{Pre-maximum spectroscopic diversity of hydrogen-poor superluminous supernovae}

\begin{abstract}

We search for the reasons behind the spectroscopic diversity of hydrogen-poor superluminous supernovae (SLSNe-I)  
in the pre-maximum phase. Our analysis is a continuation of the paper of \citet{ktr21},
who disclosed two new subtypes of SLSNe-I characterized by the presence/absence
of a W-shaped absorption feature in their pre-maximum spectra between 4000 and 5000 \AA (called Type~W and Type~15bn, respectively).
However, the physical cause of this bimodality is still uncertain. Here we present pre-maximum spectral synthesis
of 27 SLSNe-I with special attention to the photospheric temperature ($T_{ \rm phot}$) and velocity ($v_{ \rm phot}$) evolution.
We find that a $T_{\rm phot}$ limit of 12000~K separates the Type~W and Type~15bn SLSNe-I: Type~W objects tend to show
$T_{\rm phot}\geq$12000~K, while Type~15bn ones have $T_{\rm phot} \leq$12000~K.
This is consistent with the chemical composition of the studied objects.
Another difference between these groups may be found in their ejecta geometry: Type W SLSNe-I may show null-polarization,
implying spherical symmetry, while the polarization of Type 15bn objects may increase in time. This
suggests a two-component model with a spherical outer carbon-oxygen layer and an asymmetric inner layer containing heavier ions.
 The two subgroups may have different light curve evolution as well, since 6 Type~W objects show early bumps,
unlike Type 15bn SLSNe-I.
 This feature, however, needs further study, as it is based on only  a few objects at present.

\end{abstract}


\section{Introduction}\label{sec:intro}

Superluminous supernovae (SLSNe) were disclosed near the start of the 21th century, as spatially rare, but extremely bright stellar explosions usually occurring in dwarf galaxies having high specific star-formation rate and low metallicity \citep[e.g.][]{chen13, Lunnan13, Lunnan14, Leloudas15, Angus16, japelj16, perley16, chen17,Schulze18,hatsu18,nicholl21}. Their total radiated energy exceed ${\sim}10^{51}\, {\rm erg}$, leading to an extremely high absolute brightness  ($M < -21$) in the optical \citep[e.g.][]{quimby11,galyam12,nicholl15,galyam19,nicholl21}. However, this definition is not absolute: some lower luminosity objects having $\sim$-19.8 magnitudes at peak were classified as SLSNe as well by e.g. \citet{decia18,quimby18,galyam19}.

SLSNe can be divided into two main subtypes empirically, similarly to the traditional supernovae (SNe): hydrogen-poor (SLSN-I) and hydrogen-rich (SLSN-II) \citep{BW17}. The physical distinction is based on whether the progenitor star was stripped by binary interaction or pre-explosion stellar winds, or preserved its H-envelope until the moment of the explosion. 

SLSNe-II can be separated into two distinct groups. Type IIn SLSNe-I have a luminosity evolution powered by the vehement interaction with a massive, surrounding circumstellar medium (CSM) indicated by strong and narrow hydrogen emission lines in the spectra \citep[e.g.][]{smith07,drake10,manos11,benetti14}. On the other hand, "normal" SLSNe-II show ostensibly no interaction with the CSM and show broader lines in their spectra \citep[e.g.][]{miller09,gezari09,inserra18}.

In this study, we focus on the H-poor type of SLSN events that have a larger number of known examples compared to SLSN-II objects. SLSNe-I can be divided into two additional subtypes as well. Regarding photometry, fast-evolving (F) objects show a rise time from the explosion to the maximum of averagely ~30 days, while slow evolving (S) events tend to have  an average rise time  $\sim$50 days \citep{inserra17,quimby18,inserra19}. The bimodality is present spectroscopically as well: fast evolving SLSNe-I show larger photospheric velocity gradients, while slow evolving objects have low, or non-existent steepness in the photospheric velocity evolution \citep[see e.g.][]{inserra18,ktr21}.

Recently \citet{ktr21} found that the time scales of photometric and spectroscopic evolution are not necessarily the same: spectroscopically fast evolving events were revealed to be fast-evolving in terms of photometry as well, however, some spectroscopically slow  SLSNe-I may show fast LC rise-times. Thus one cannot say simply that a SLSN is a "Fast" or "Slow" evolving event, but may put these categories together with the "photometrically" and "spectroscopically" attributes.

Rather than using a somewhat arbitrary brightness cut at -21 mag, SLSNe-I now are instead identified by their unique optical spectra showing a steep blue continuum together with a series of broad and blended absorption features.  According to e.g. \citet{quimby11,mazzali16,liu17,quimby18}, the lines of O II appear in the spectra together with the weaker features of Si II \citep{inserra13}, Fe III \citep{lelo12,nicholl13} and C II \citep{nicholl16}.  The spectra of SLSNe-I show a great resemblance to the spectra of normal Type Ic or broad-line Ic SNe, but with a $\sim$30 days delay, which means that the $\sim$ +30 $--$ +50 days phase spectrum of a SLSN-I shows similar features to SN Ic/BL-Ic spectrum near the maximum \citep[see e.g.][]{pastor10,inserra13,nicholl13,blanchard19}. Therefore \citet{nicholl21} suggested that SLSNe may be referred as "ultra-hot" Type Ic SNe with a reasonably slower evolution compared to SNe Ic that are caused by their larger peak luminosity. The higher temperatures of SLSNe can explain the extremely blue continuum with peculiar features in the early pre-maximum phase, which can be used to distinguish SLSNe-I from normal SNe Ic or BL-Ic. One of these early-phase features present only in the case of SLSNe-I is a peculiar W-like absorption blend appearing between 3900 and 4500 \AA.  It is usually identified as a blend of singly ionized oxygen lines, and is believed to play a major role in the spectrum formation of  most SLSNe-I in the pre-maximum phases \citep[e.g.][]{quimby11,mazzali16,liu17}. Alternatively, this feature can be modeled using the mixture of different ions, e.g. O III and C III \citep{quimby07, dessart19, galyam19b, ktr20-2}. However, \citet{ktr21} revealed that this W-shaped feature is missing from the spectra of some SLSNe-I that were found to be spectroscopically similar to SN~2015bn. Thus, two new subgroups of SLSNe-I were created regarding their pre-maximum spectra by the presence or absence of the W-shaped absorption blend: Type W and Type 15bn  SLSNe-I, respectively. 

According to \citet{ktr21}, Type W SLSNe-I are represented in both Fast and Slow subgroups regarding photometry and spectroscopy as well, while SN~2015bn-like SNe are present only in the spectroscopically slow evolving group with LC rise-times ranging between a few 10 days to $\sim$150 days.

The main goal of this study is to search for physical differences between Type W and Type 15bn SLSNe-I. As we do not possess a large set of data to carry out a statistical investigation, this paper shows a comparative pre-maximum spectral analysis of 27 SLSNe-I. We investigate here only the pre-maximum phase, as the characteristic W-shaped absorption feature is present before the moment of the maximum. Post-maximum analysis of the sample is planned in a future paper. 

In Section \ref{sec:data}, we describe the data we utilized and give details of our main motivations to do this study. In Section \ref{sec:model}, we show the spectral synthesis and ion identification of the available spectra, separately for Type W and Type 15bn SLSNe-I. In Section \ref{sec:disc} we discuss our findings regarding the spectroscopic differences between the two groups in terms of chemical composition, photospheric temperature and velocity evolution and explosion geometry.  Section \ref{sec:sum} summarizes our results.

\section{Motivation and data}\label{sec:data}

In some respects, this study is a continuation of the recent paper of \citet{ktr21}, and therefore we utilize the same sample of 27  SLSNe-I with photometric and spectroscopic data downloaded from the Open Supernova Catalog (OSC)\footnote{https://sne.space/} \citep{guill17}. The sample selection method and the basic data of the objects are detailed in Section 2 and Table 1 of \citet{ktr21} , where the original references  are presented for each studied object.  It is important to note that the sample analyzed in \citet{ktr21} contained 28 members. In this paper we examine only 27 of them, since SN2017faf was removed from the sample due to the uncertain origin of this transient. First it was classified as a SLSN-I due to the W-shaped O II absorption line in its pre-maximum spectra \citep{kilpatrick17}, however, in the early post-maximum phase ($\sim$2 days after maximum) H$_{\alpha}$,$H_{\beta}$ and He I $\lambda$5876 were identified in the spectra, and therefore SN~2017faf was re-classified as a Type II SN \citep[see][]{pastor17}.

In the following, we describe our motivation to examine these objects further. 
In the \citet{ktr21} paper, the main goal was to search for possible correlations between the light curve (LC) evolution time scales, and the ejecta masses of the examined objects, for which we utilized the following two equations: 
\begin{align}
M_{\rm ej} ~=~ 4 \pi {c \over \kappa} v_{ \rm ph} t_{\rm rise}^2   &&
M_{\rm ej} ~=~ {{\beta c} \over {2 \kappa}} v_{\rm SN} t_m^2.
\end{align}

Here, $M_{\rm ej}$ is the ejecta mass of the objects, while $v_{\rm ph}$ is the velocity of the photosphere measured pre- or near-maximum, $t_{\rm rise}$ is the rise-time of the LC from the explosion to the maximum, $\beta \approx 13.8$ is an integration constant, slightly depending on the ejecta density profile,  $v_{\rm SN}$ is a scaling velocity of the optically thick ejecta, and $t_m$ is the ``mean light curve timescale''.  The $t_{\rm rise}$ of each event was calculated from the dates of explosion and maximum light estimated by the MOSFit fittings available in the Open Supernova Catalogue. Here, it is seen that photospheric velocity values near maximum are required to infer the ejecta masses of the objects.

\begin{figure*}
\centering
\includegraphics[width=12cm]{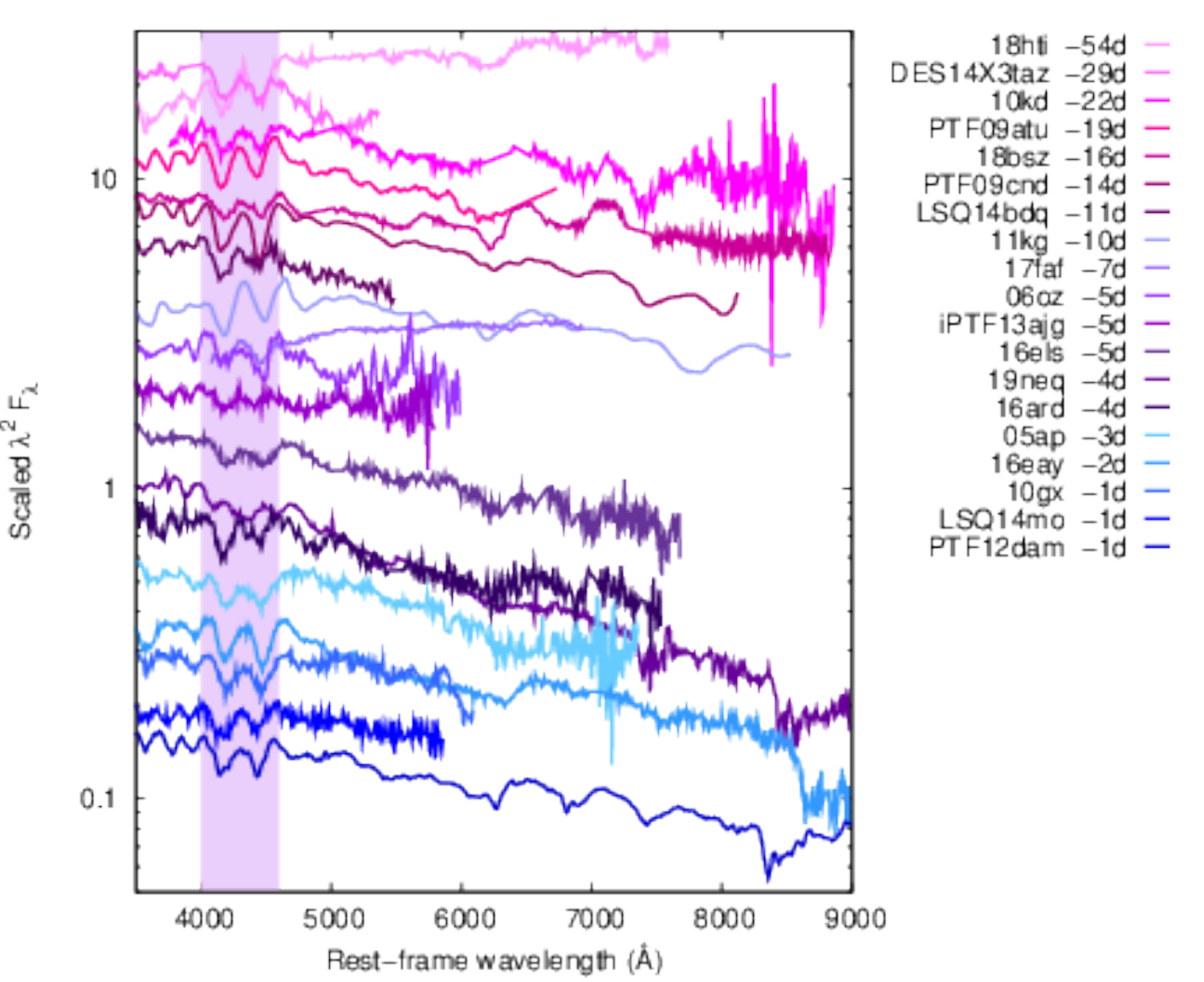}
\includegraphics[width=12cm]{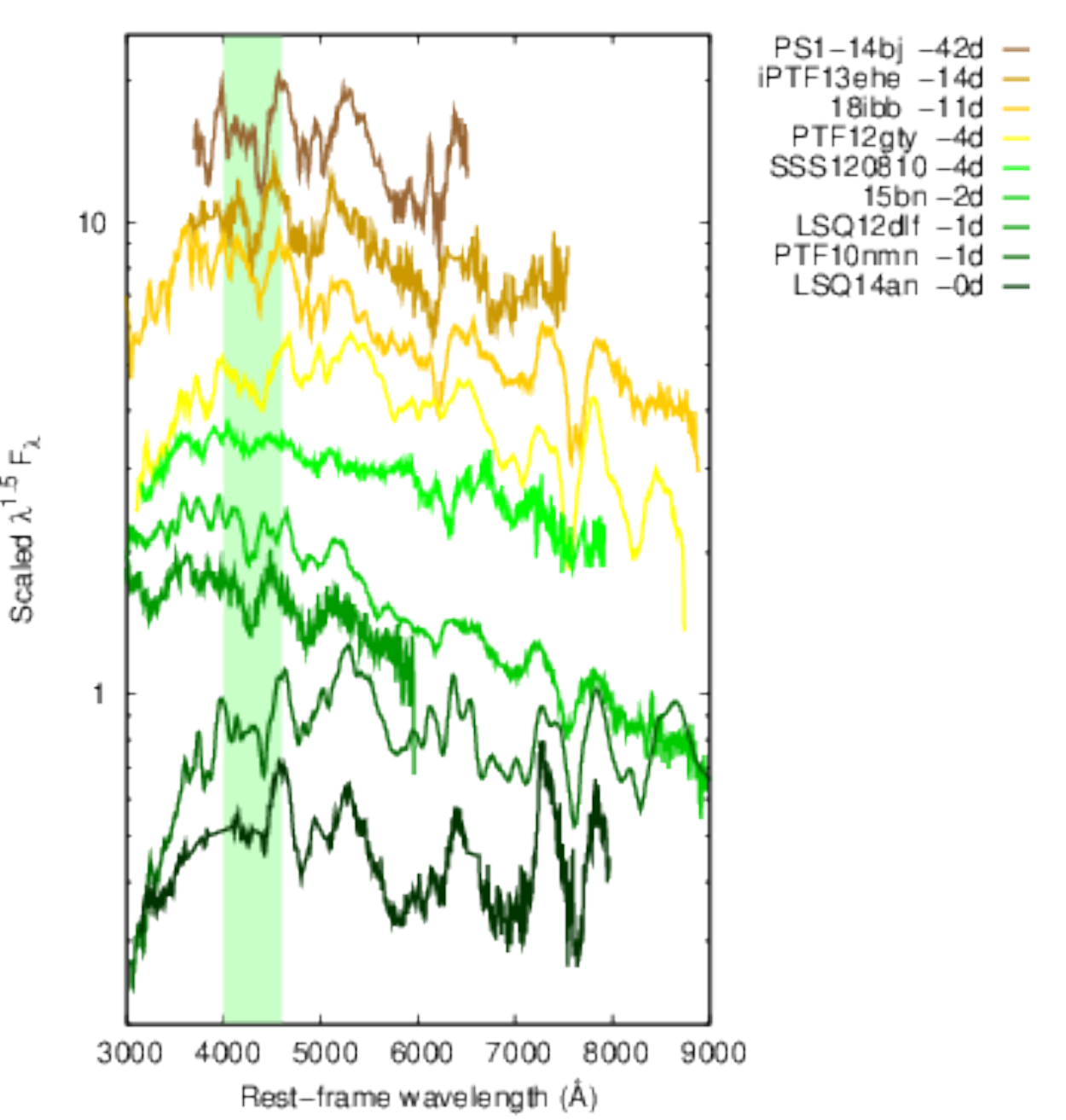}
\caption{Top panel: Pre-maximum spectra of Type W SLSNe-I sorted by phase. The W-shaped absorption feature is apparent near 4000-5000 \AA\ in these spectra, which region is highlighted with purple. Bottom panel: Pre-maximum spectra of Type 15bn SLSNe-I sorted by phase. The wavelength range, where the W-shaped feature is expected, is shaded with green. It is seen that such absorption blend is missing from all spectra of Type 15bn objects. The narrow emission lines of the host galaxy, as well as the telluric absorption lines, were removed from all spectra before plotting.}
\label{fig:classes}
\end{figure*}

In order to estimate the photospheric velocities in a fast, but reliable way, a method combining the spectrum modeling and the cross-correlation technique was developed \citep{modjaz05,ktr21}, and as a secondary product of these calculations, two new subtypes of SLSNe-I were disclosed. The method to infer the $v_{\rm phot}$ is based on some previous findings that suggest that the pre-maximum spectra of all  SLSNe-I are dominated by a W-shaped  O II feature near 4000 \AA\ \citep[e.g.][]{quimby11,mazzali16,liu17}. Therefore a series of SYN++ \citep{thomas11} models with different photospheric velocity values were built at the suspected wavelength of the W-shaped feature and then cross-correlated with the observed spectra. However, in  case of the 1/3 of the sample, this method did not work well and led to physically senseless $v_{\rm phot}$ values. The cause of the discrepancy was revealed to be in the observed spectra: 9 objects of the sample actually did not show the expected W-shaped feature. After collecting those spectra, they were found to be similar to each other, and they were named by a representative and well-observed example object, SN~2015bn. In this manner, two new subgroups of  SLSNe-I were separated by the presence/absence of the W-shaped features in the pre-maximum spectrum: Type W objects that show the W-shaped feature, and Type 15bn objects, where the W-shaped absorption lines are missing. 

The distinction between these two groups is presented in Figure \ref{fig:classes}. In the top panel, some pre-maximum spectra of Type W SLSNe-I are shown sorted by ascending phase. All plotted spectra were normalized to the flux at 6000 \AA, and corrected for redshift and Milky Way extinction. The strong and narrow emission lines and the telluric absorption features were removed for clarifying as well, as was presented by \citet{quimby18} as well. The semi-transparent purple shading highlights the wavelength region, where the W-shaped absorption blend is present. The bottom panel of Figure \ref{fig:classes} shows the pre-maximum spectra of the objects that belong  to Type 15bn subgroup. It is seen that in the shaded region, the W-shaped lines, which were clearly present in the top-panel spectra, are missing here, making the difference between Type W and Type 15bn  SLSNe-I apparent.

\citet{ktr21} examined the spectroscopic- and photometric evolution timescales of the 27 SLSNe-I in their sample, and found that Type W and Type 15bn  SLSNe-I are evolving differently. 
All Type 15bn  SLSNe-I  in the sample showed nearly constant photospheric velocity evolution between the pre-maximum phases and +30 days phase after the moment of the maximum light, while Type W  SLSNe-I  were represented in the low- and the high velocity gradient group as well.  A new method was developed by \citet{ktr21} for the classification of the objects into the Fast or Slow evolving subgroup. Following the methodology of \citet{inserra18}, the photospheric velocity gradients between the moment of the maximum and $\sim$30 days after were calculated for 9 objects in the sample that had available spectroscopic data near +30 days phase after the moment of the maximum. From this calculation, it was found that the larger the velocity at maximum, the steeper its decline in the later phases (see Figure 11 of \citet{ktr21}, and \ref{tab:osszefoglalo}), and therefore one spectrum taken near maximum is sufficient to decide if a SLSN-I belongs to the Fast or the Slow evolving subgroup.  Spectroscopically Fast evolving SLSNe-I were defined to have  $v_{\rm phot} \geq 20000$ km s$^{-1}$ near maximum, while the objects showing  $v_{\rm phot} \leq 16000$ km s$^{-1}$ were classified as spectroscopically Slow evolving. \citet{inserra18} showed a correlation between the velocity gradient and the velocity earlier as well, and found that for the SLSNe-I with a velocity gradient lower than $\sim$100 km s$^{-1}$ day$^{-1}$, the correlation does not hold anymore (see Figure 7 of \citet{inserra18}). This means that we cannot decide unambiguously that the object is Fast or Slow evolving from the velocity gradient, when its value is lower than $\sim$100 km s$^{-1}$ day$^{-1}$. However, from the photospheric velocity at maximum itself gives a clue for the fastness/slowness of the evolution, and therefore, instead of velocity gradients, single values of photospheric velocities near maximum may be used for classification.

But what can be the physical distinction of such groups? Are Type W and Type 15bn  SLSNe-I  different in their photospheric temperatures? Is the cause of the missing O II line lie in this phenomenon, or some other effects, maybe the symmetry/asymmetry of the ejecta? In order to search for answers to these questions, and examine the photospheric temperature- and velocity evolution of these SLSNe-I, all spectra in the sample were modeled utilizing the SYN++ code. It is important to note here that when examining the effect of the temperature at the photosphere on the observed spectra, we assume that the relative fluxes of each object downloaded from the Open Supernova Catalog, and then corrected for Milky-Way extinction, are reasonable.

This study focuses only on the pre-maximum phase, as the presence/absence of the W-shaped feature, so the main and most visible difference between Type W and Type 15bn  SLSNe-I  can only be observed before the maximum \citep{quimby18}.  It is an interesting question, whether Type W and Type 15bn  SLSNe-I  are showing visible differences after the maximum as well, however, the post-maximum study of these objects is the topic of a planned future paper.

\section{Pre-maximum spectrum synthesis}\label{sec:model}

In this section, we discuss the pre-maximum spectrum modeling of the 27 SLSNe-I (19 Type W and 9 Type 15bn) in our sample, for which we utilized the code named SYN++. The velocity ($v_{\rm phot}$ and the temperature ($T_{\rm phot}$) at the photosphere are important parameters of the synthesized spectra together with the optical depth ($\tau$), the feature width velocity ($aux$), and the excitation temperature ($T_{\rm exc}$) of each identified ion. All spectra were continuum-normalized using the fitted Planck function after the modeling in order to make the spectral features more visible.

\subsection{Possibilities and constraints of the spectrum synthesis with SYN++}\label{subsec:syn}

The SYN++ code works with some basic assumptions, which are the following \citep{thomas11, BW17}:
\begin{itemize}
    \item Spherical symmetry
    \item Homologous expansion
    \item Opaque photosphere emitting blackbody radiation
    \item Spectral features form in a semi-transparent shell above the photosphere
    \item The line forming mechanism is purely resonant scattering
    \item The velocity gradient is so large that the places showing the same Doppler-shift  are located on a thin plane perpendicular to the line of sight (Sobolev-approximation)
\end{itemize}

A code based on these  assumptions certainly has numerous caveats. One of the most important limitations is the assumption of the blackbody radiation of the photosphere, which implies Local Thermodynamic Equilibrium (LTE) conditions that rarely  happen in reality. However, according to some previous studies \citep[e.g.][]{BW17}, the LTE conditions  may be applicable in the case of supernovae with larger ejecta masses, at least around maximum light. Therefore, assuming LTE is not necessarily valid for normal Type Ia SNe usually showing small ejecta masses, but can be utilized while modeling the pre- and near maximum spectra of Type II SNe and SLSNe having larger  ejecta masses. 

In this study, we focus on the spectral synthesis of Type I SLSNe before the moment of the maximum, thus, using the SYN++ code seems to be reasonable. It is important to note, though that the pseudo-continuum of some spectra does not always follow the Planck function of the pure blackbody radiation, and show deviations, in most cases a decrease  of flux in certain wavelength regimes. This suggests that some non-LTE effects and/or continuous absorption/scattering processes may play a role in the spectrum formation of SLSNe-I as well, which are not taken into account in SYN++. However, considering all non-LTE effects that may affect the observed spectrum is beyond the scope of this study, but they should be kept in mind.

 It has to be mentioned that the SYN++ code assumes spherical symmetry, which may not be valid in the case of some studied SLSNe-I. In Section \ref{subsec:geom}, we will discuss in details the results of polarimetric observations carried out earlier, which revealed a few SLSNe-I, where \sout{there} an increase in polarization was detected. It was explained with a two-layered model consisting of a more symmetric outer layer and a less symmetric inner layer of the ejecta. One of the main consequences of the deviation from spherical symmetry is that the whole system will show a viewing angle effect that may be different in each cases. Even though it would be very important to take this effect into account, it needs a more sophisticated modeling than applied here, thus, it is beyond the scope of the present paper. It is planned as a possible future project. 

 SYN++ was found being very useful in the identification of elements/ions that form the observed spectral features.  However, element identifications  can be somewhat ambiguous as well, which is especially true for SLSNe, since these objects usually show lots of features that are broad and blended. Pre-maximum spectra of SLSNe-I tend to show a hot blue continuum with only a few absorption lines, thus, the mentioned effect is less significant compared to the near- and post-maximum phases. Since Type W SLSNe-I tend to show larger photospheric temperatures than Type 15bn  SLSNe-I  (see Section \ref{sec:disc} for detailed discussion), the line-blending of lots of  features plays a smaller role in their spectrum formation. However, it was shown previously e.g. by \citet{dessart19} and \citet{ktr20-2} that the W-shaped feature between 4000 and 5000 \AA\ can be fitted equally well not only using O II and C II lines but  with O III and C III as well. Therefore it can be concluded that apart from the strongest lines in the spectra, the ion identification has a lot of limitations. 

Another kind of ambiguity can be explained by the example of the Ca II lines. This ion usually shows two  groups of very strong features in the spectrum. However, these lines  are located in the bluest and the reddest parts of the spectrum, respectively, therefore,  they may be missing from  the observed spectra due to e.g. redshift. It is seen that not only the limitations of the models but  also those of the observed spectra have an impact on the ion identification as well. 

The ion compositions of the SYN++ models that we build are able to fit the observed features, but for the reasons mentioned above they have to be treated cautiously. Running some complex explosion and NLTE spectrum synthesis models would  probably give more reliable ejecta compositions. However, it is beyond the scope of the present paper. Spectrum synthesis with SYN++ may be the first step  in understanding the differences between Type W and Type 15bn SLSNe-I.

\subsection{Type W objects} 

\begin{figure*}
\centering
\includegraphics[width=5cm]{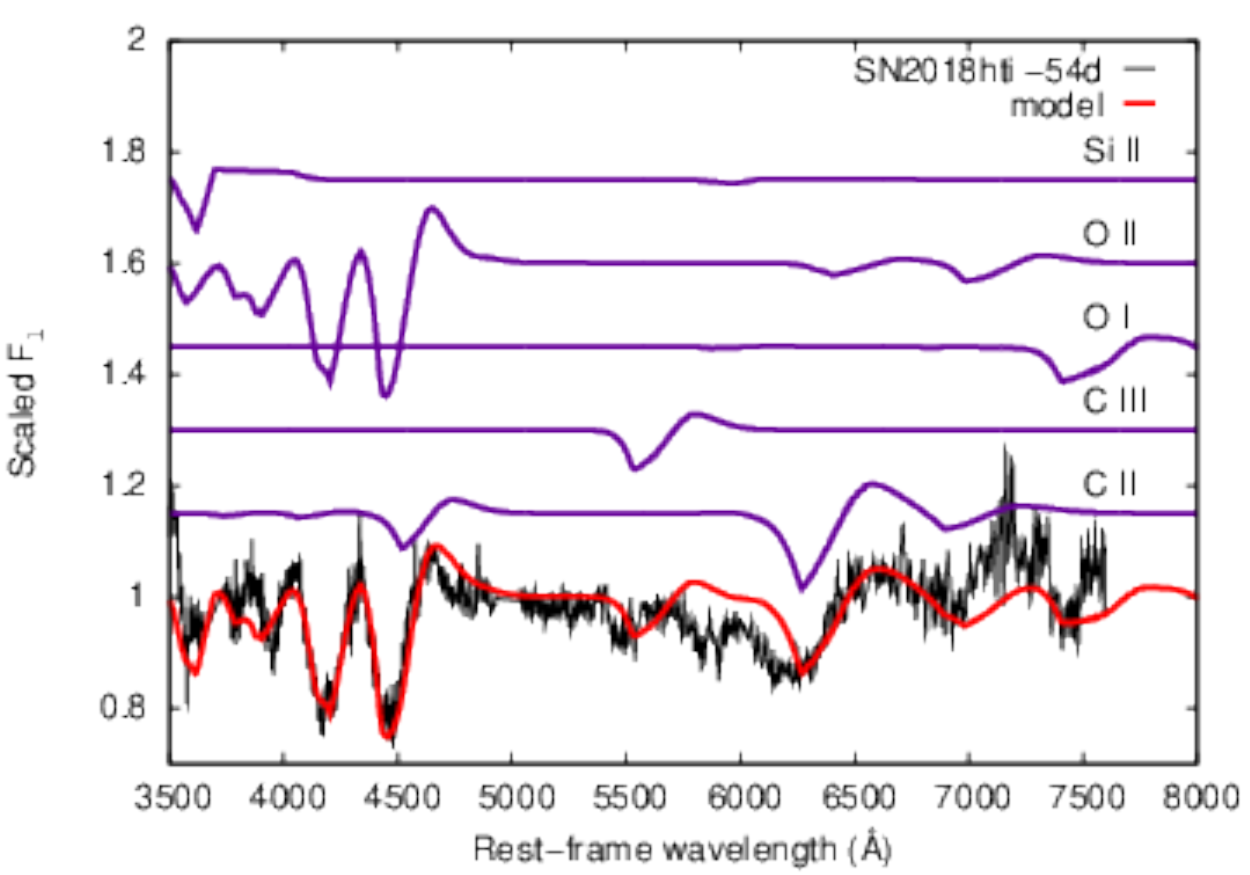}
\includegraphics[width=5cm]{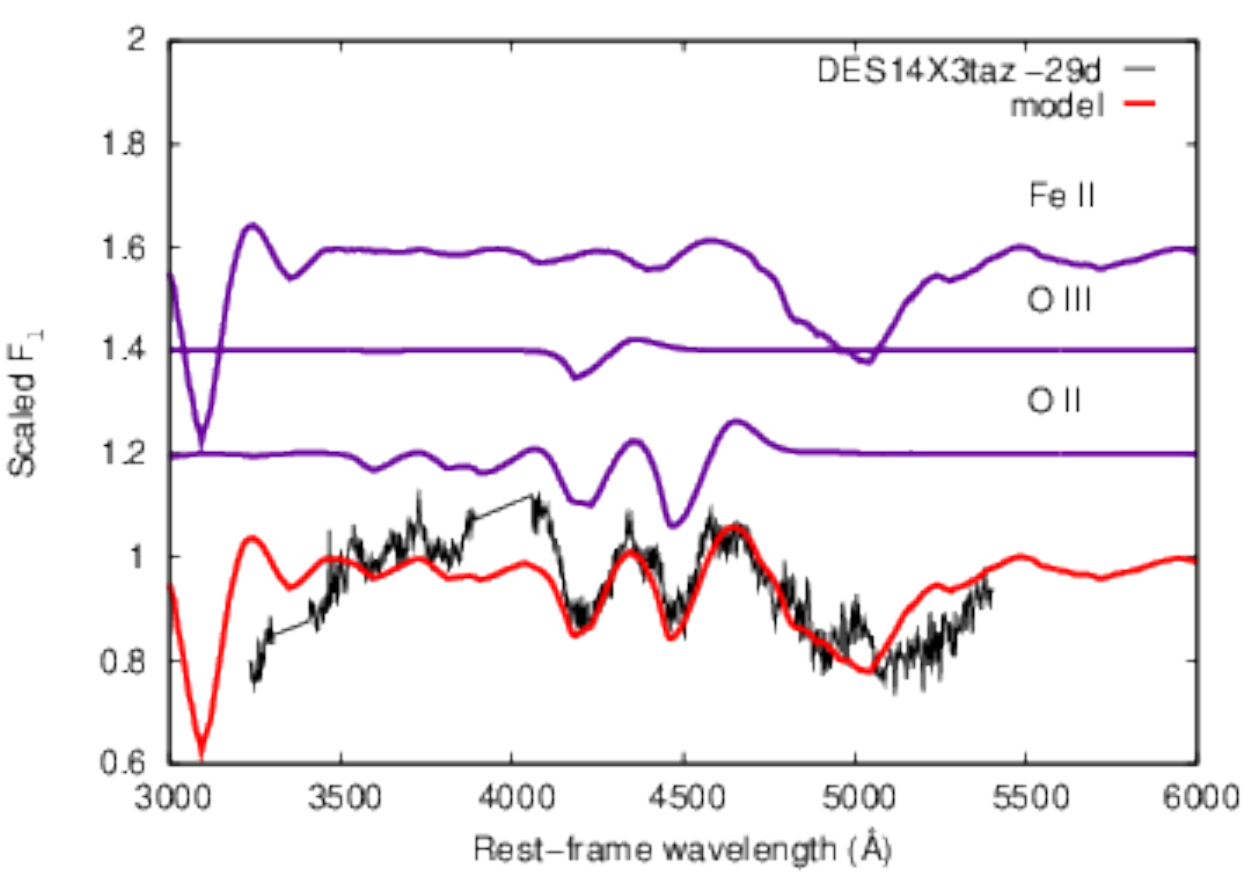}
\includegraphics[width=5cm]{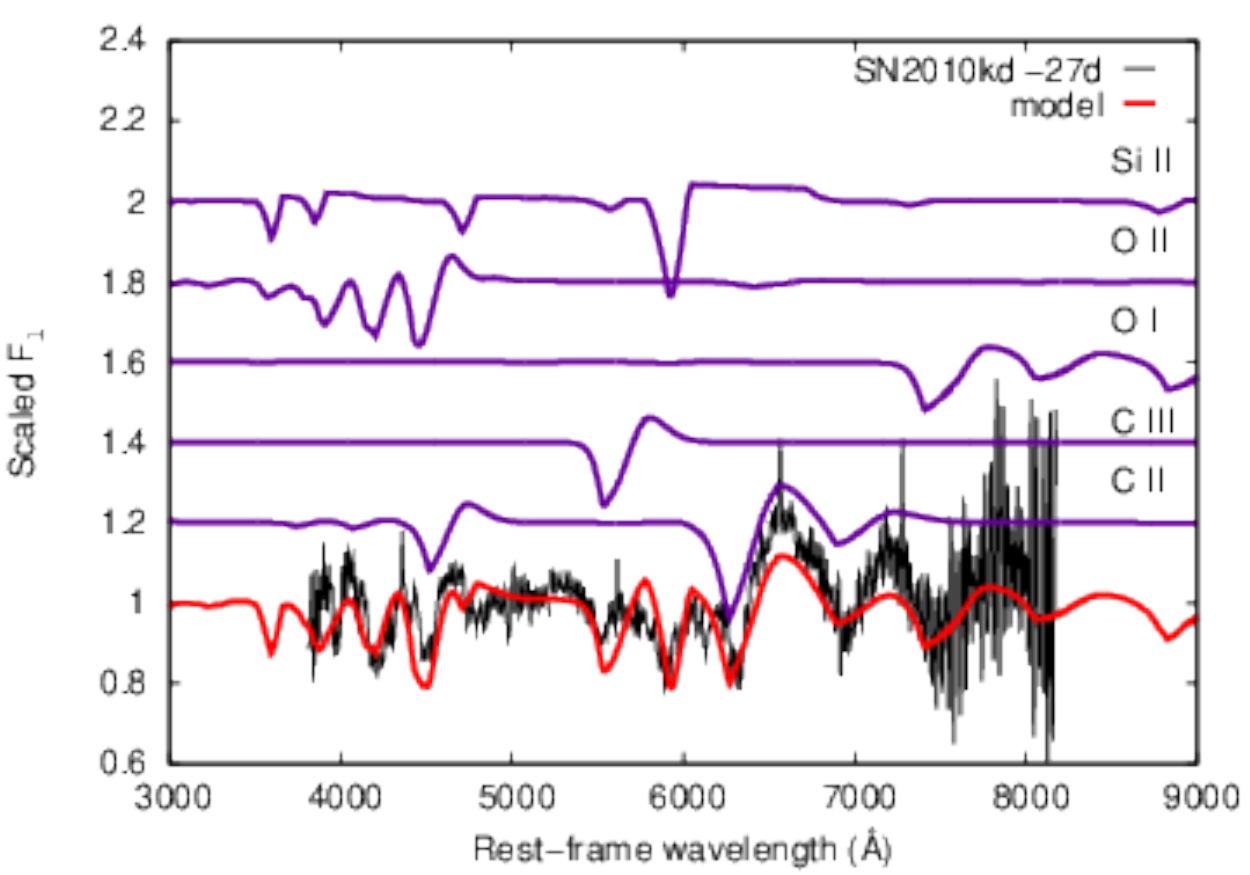}
\includegraphics[width=5cm]{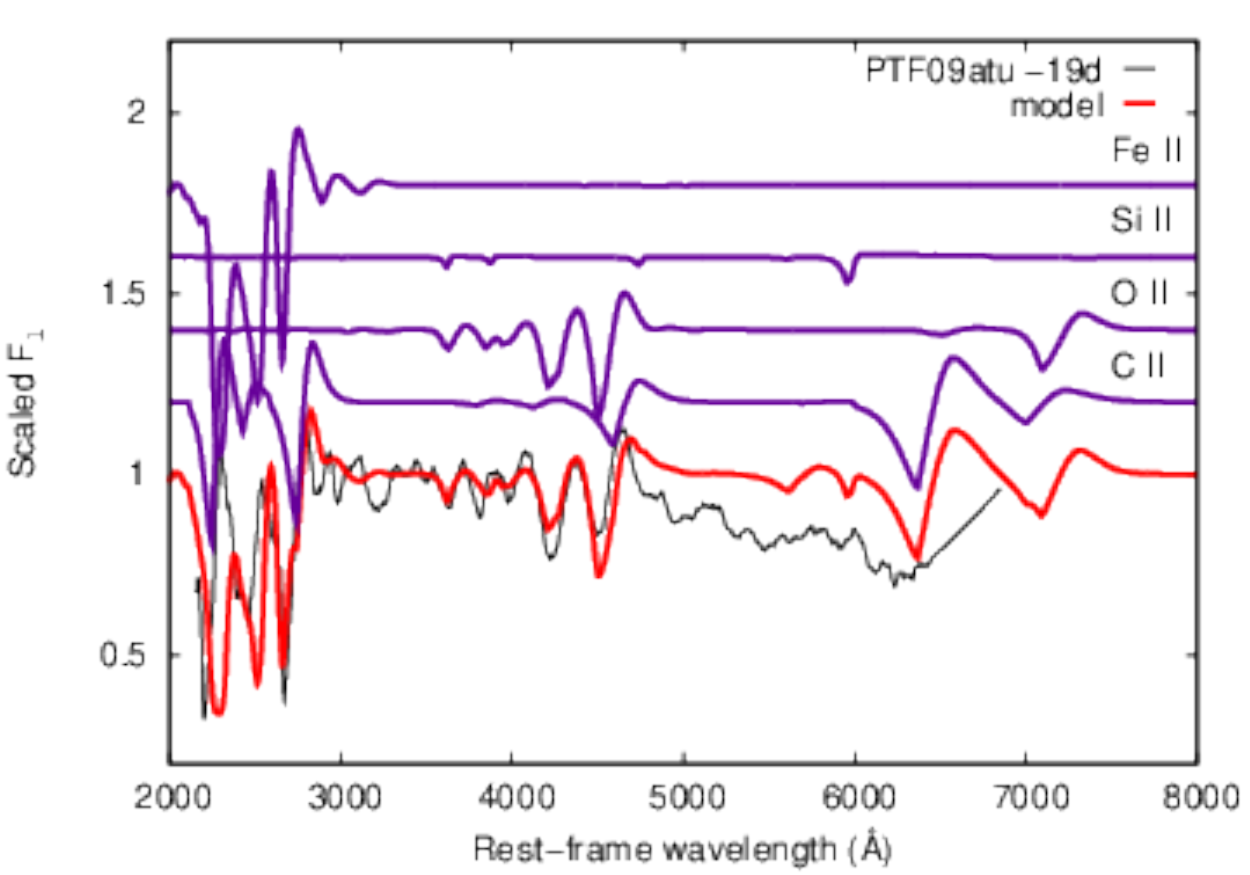}
\includegraphics[width=5cm]{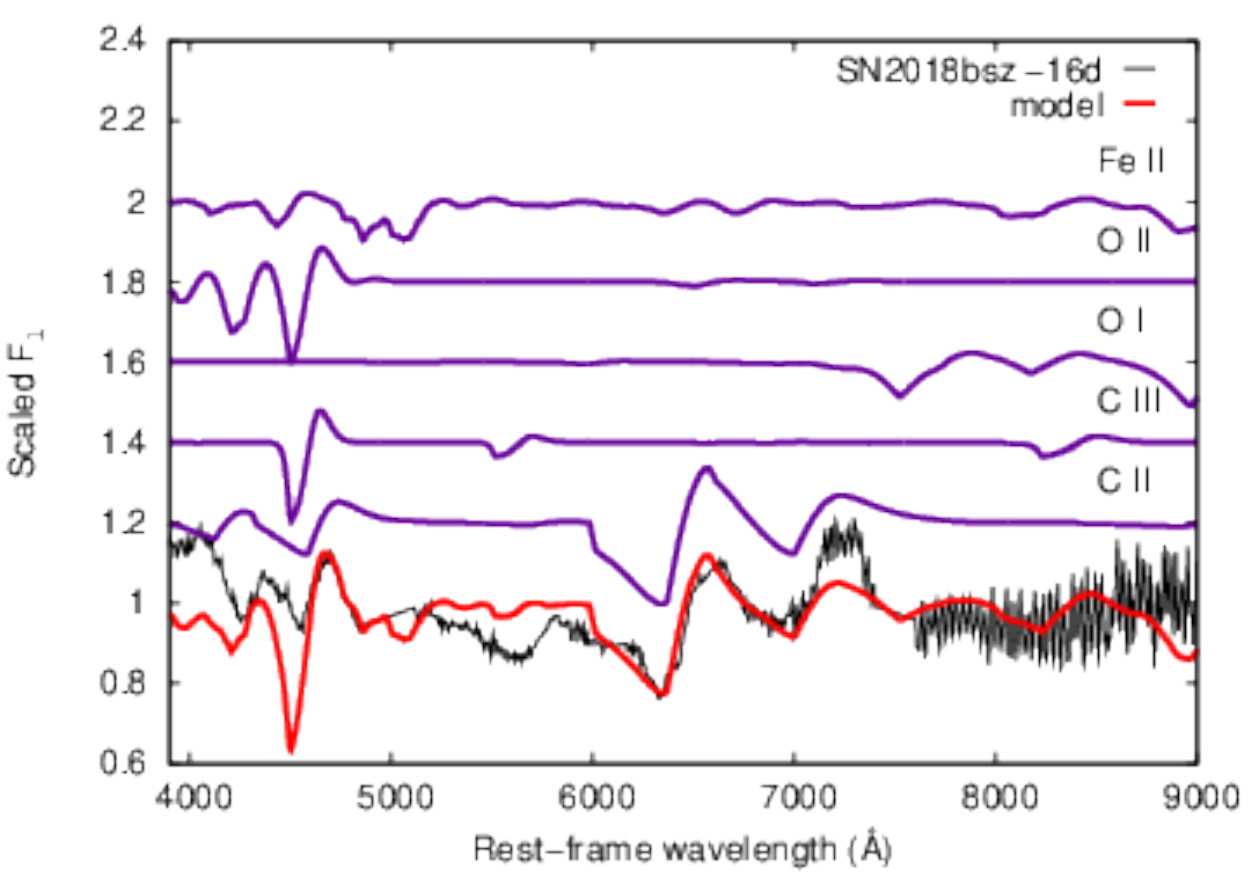}
\includegraphics[width=5cm]{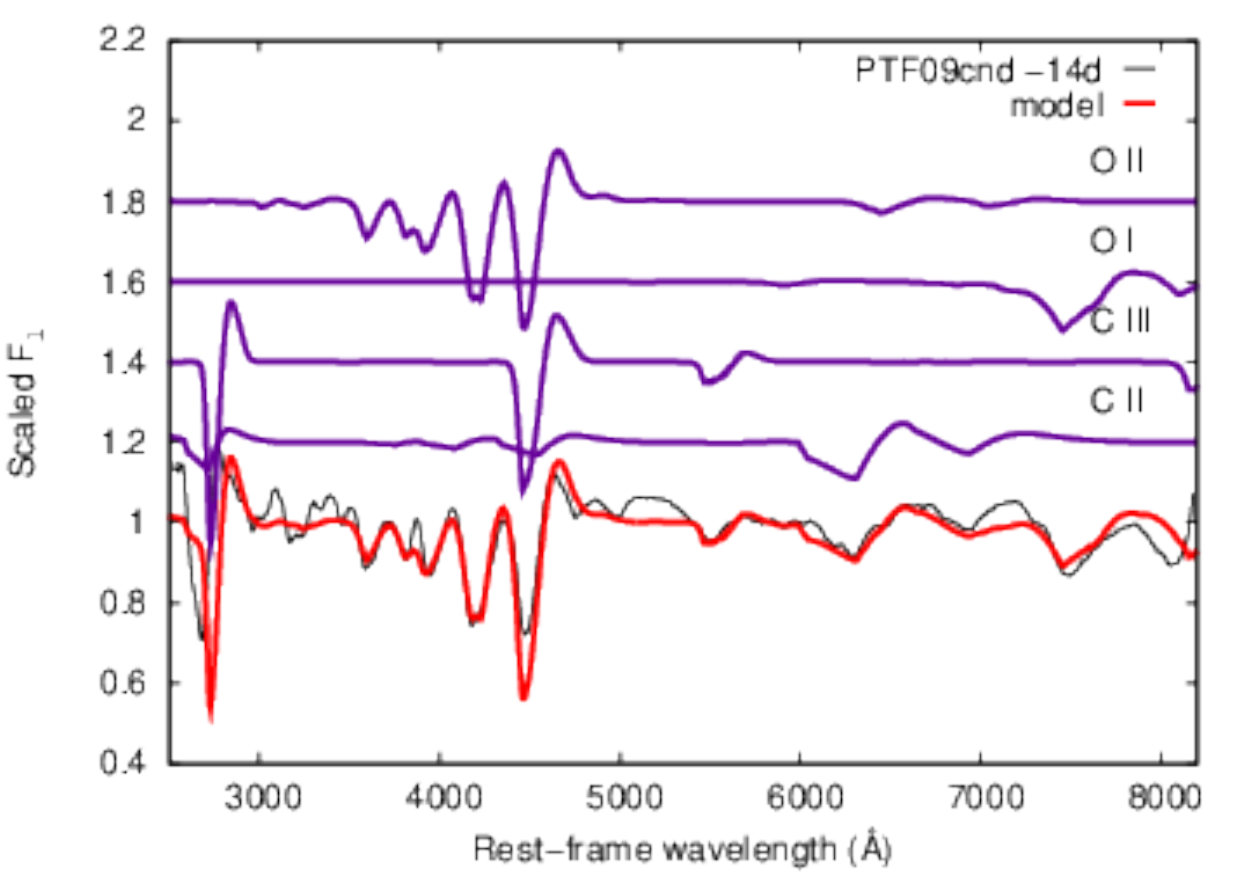}
\includegraphics[width=5cm]{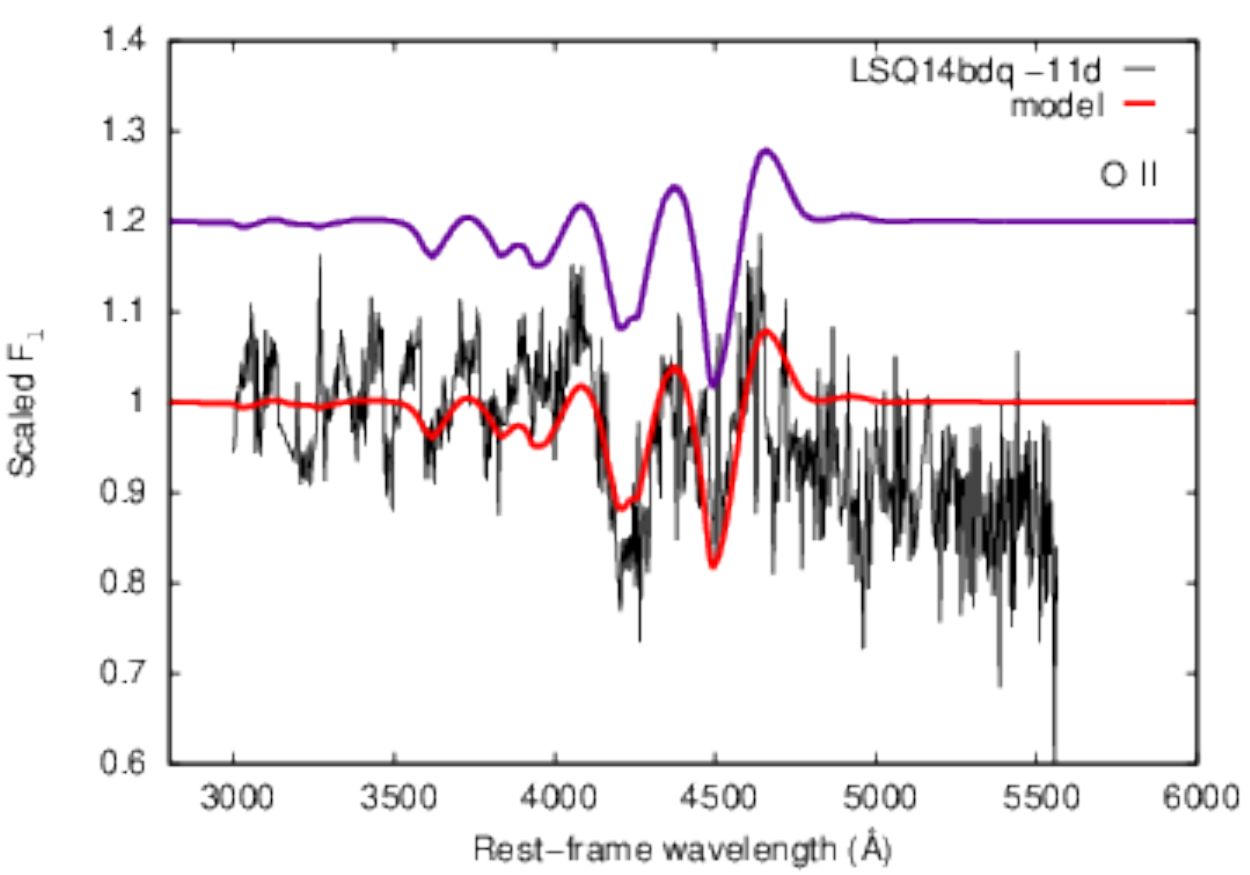}
\includegraphics[width=5cm]{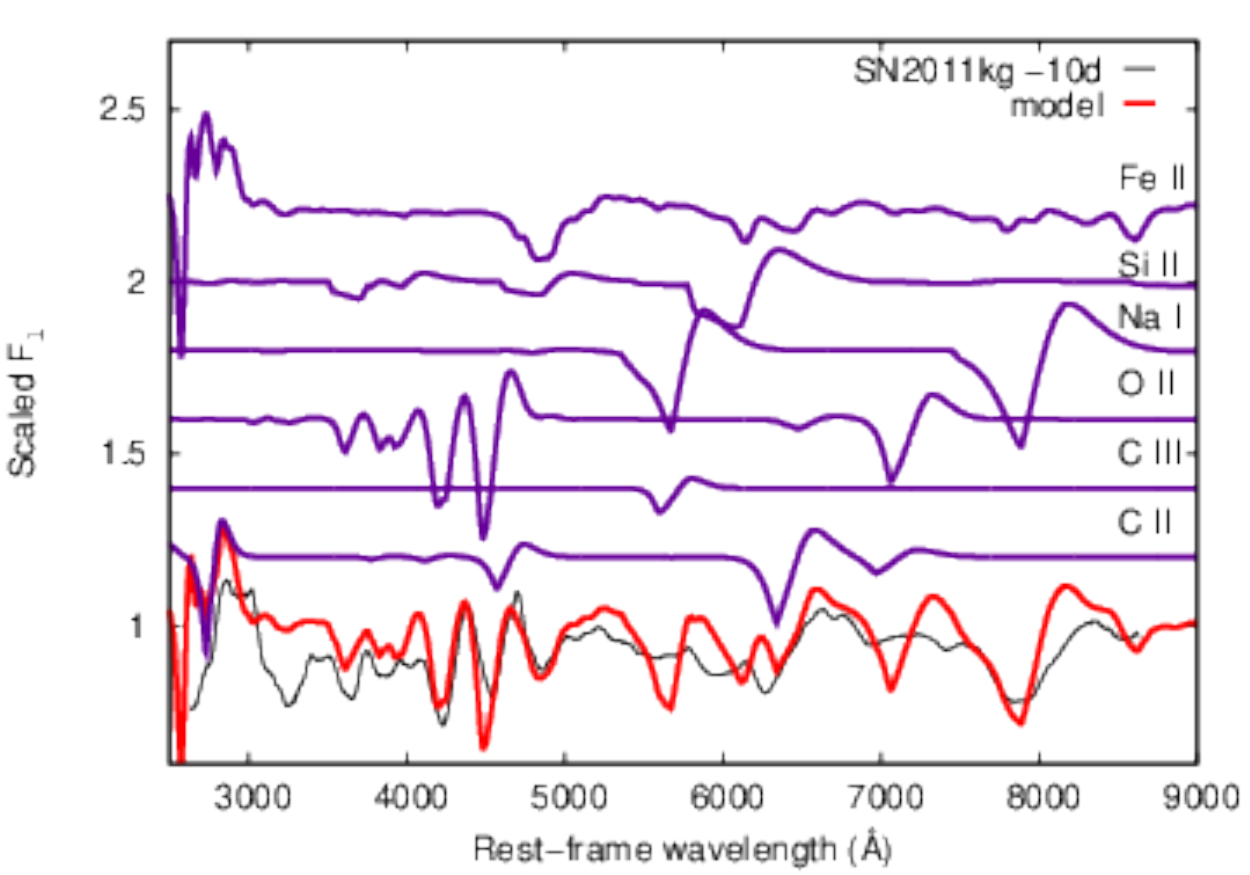}
\caption{Early pre-maximum spectra of Type W  SLSNe-I . The observed spectra are plotted with black, while red color indicates the best-fit models obtained in SYN++. The single-ion contributions to the overall model (purple lines) are shifted vertically for clarification. }
\label{fig:early_wmodels}
\end{figure*}

\begin{figure*}
\centering
\includegraphics[width=5cm]{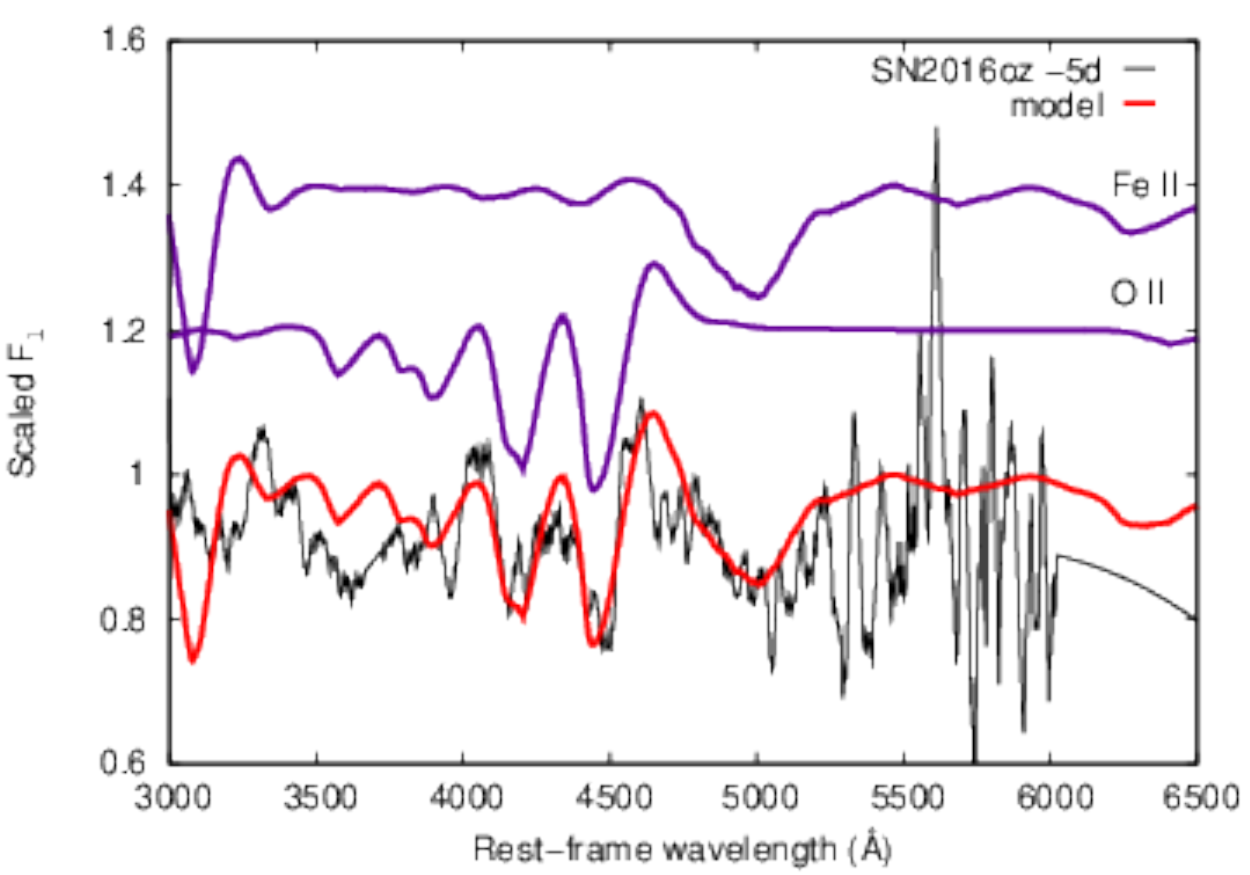}
\includegraphics[width=5cm]{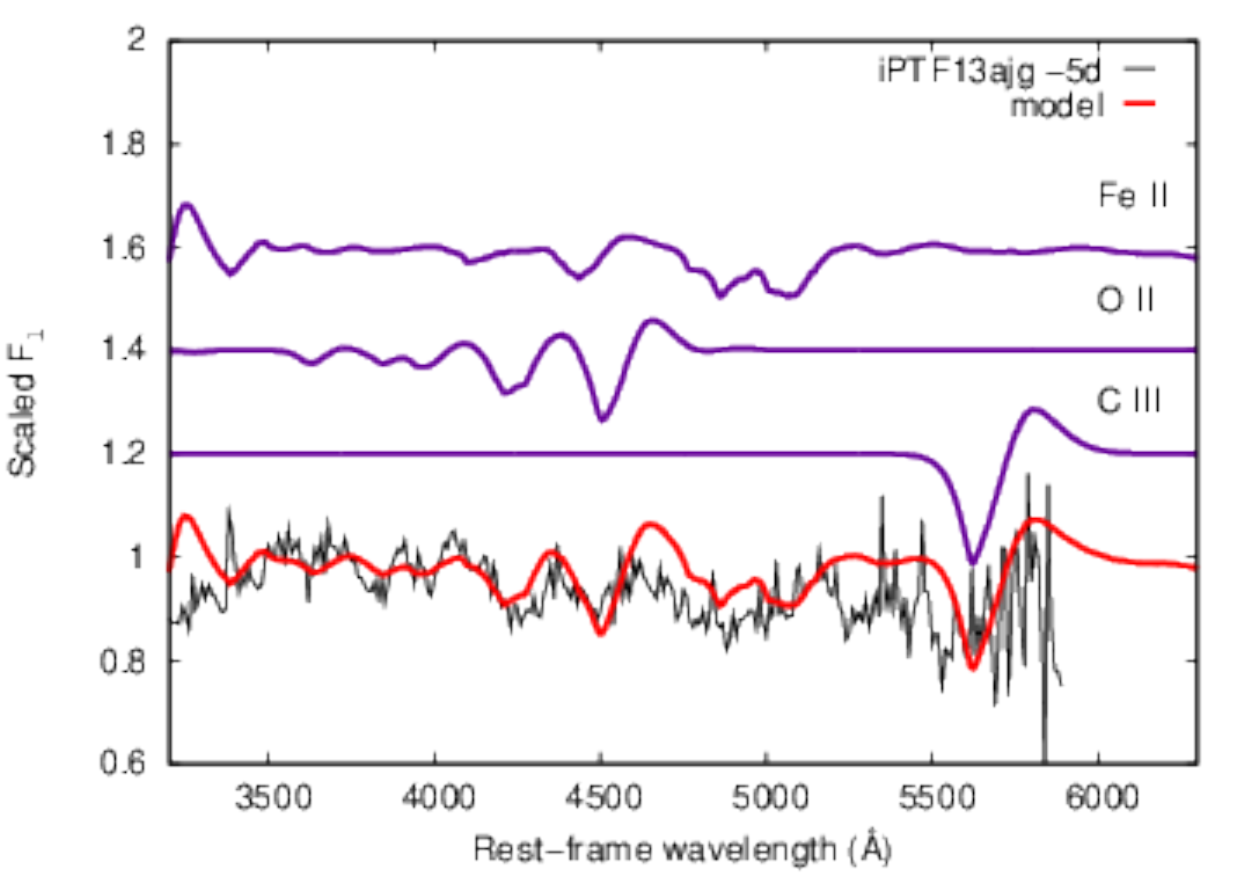}
\includegraphics[width=5cm]{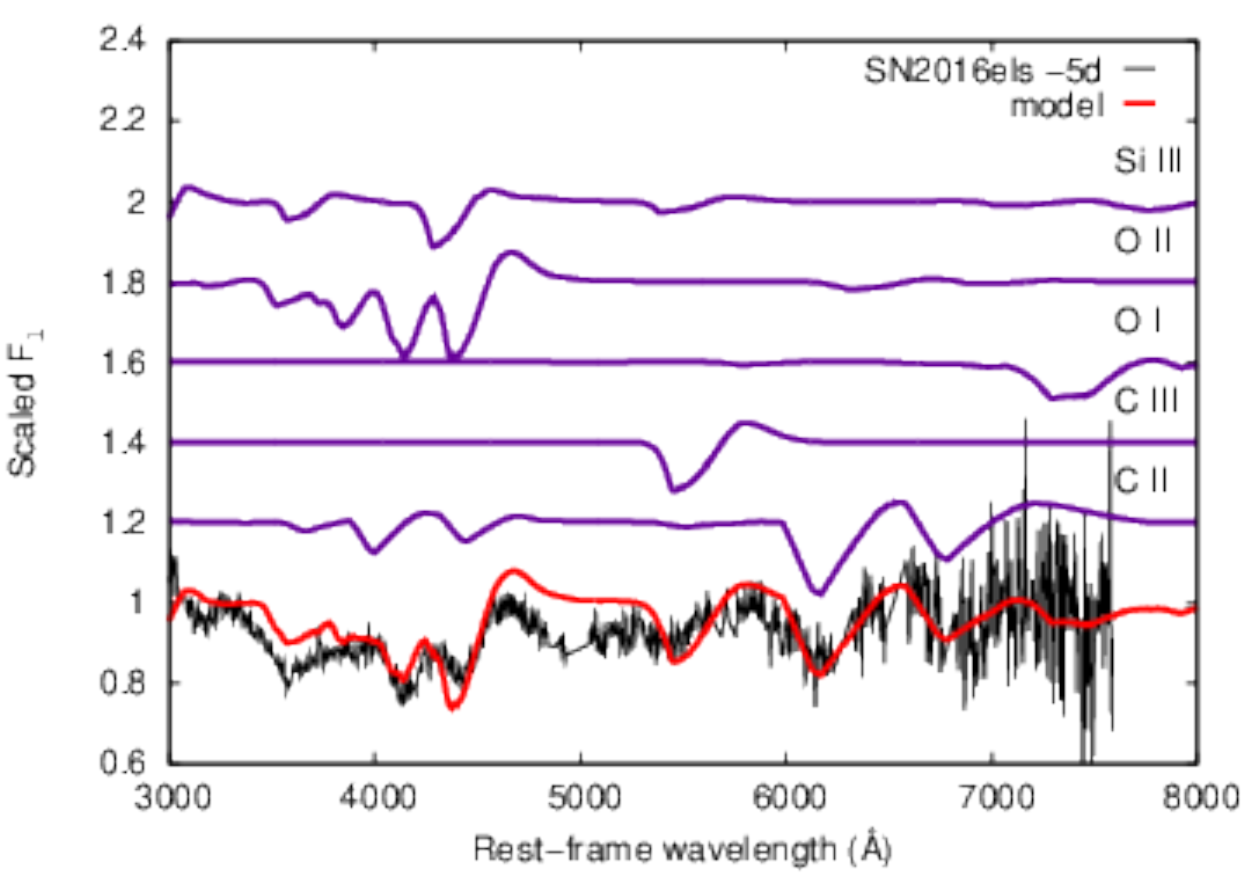}
\includegraphics[width=5cm]{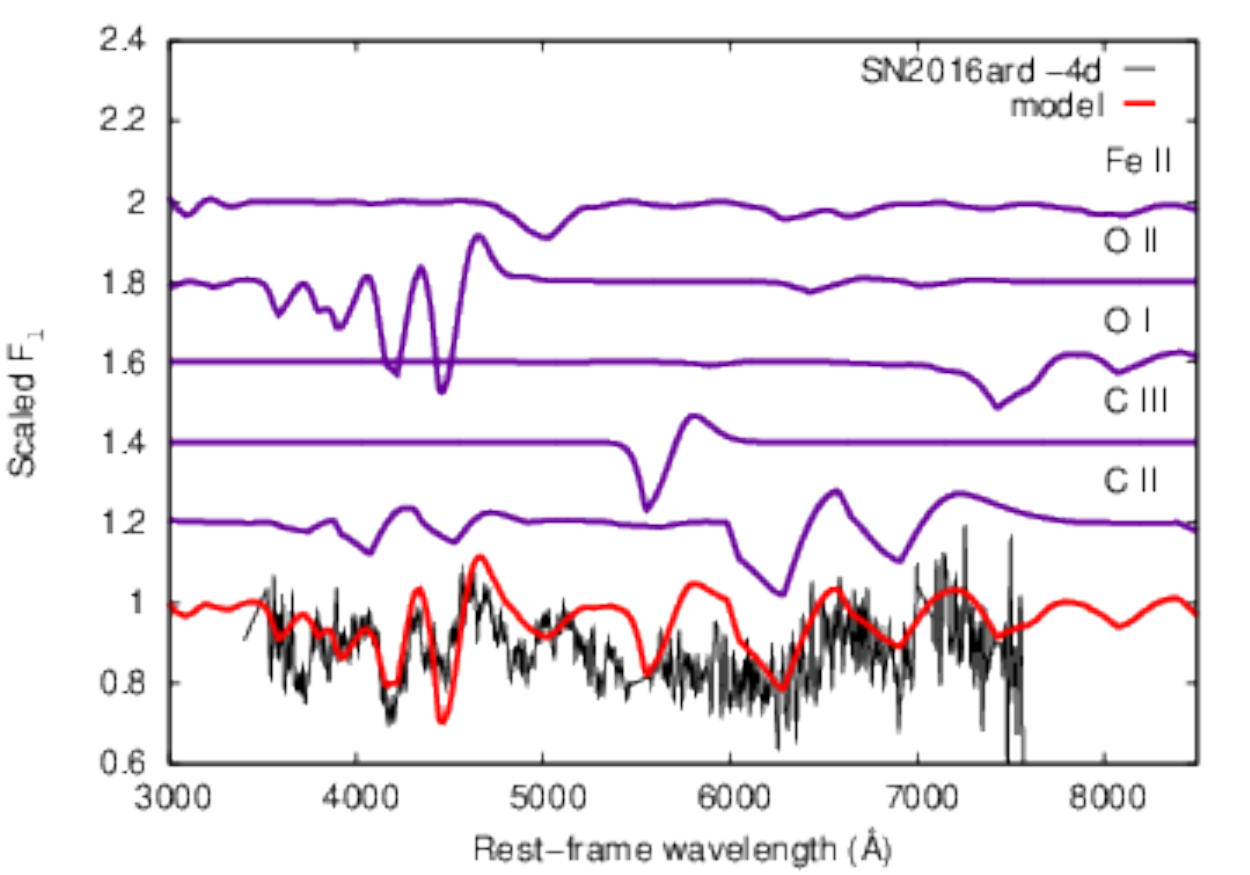}
\includegraphics[width=5cm]{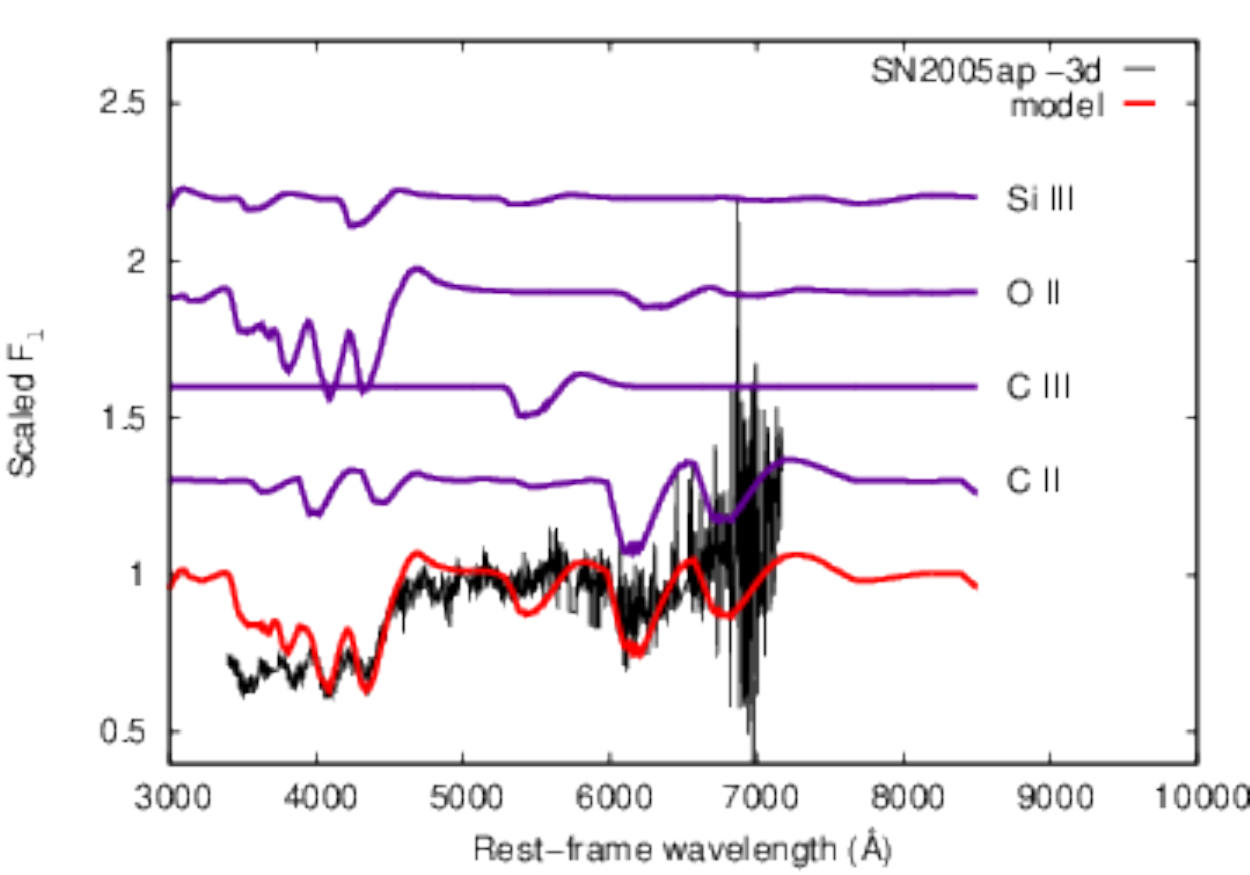}
\includegraphics[width=5cm]{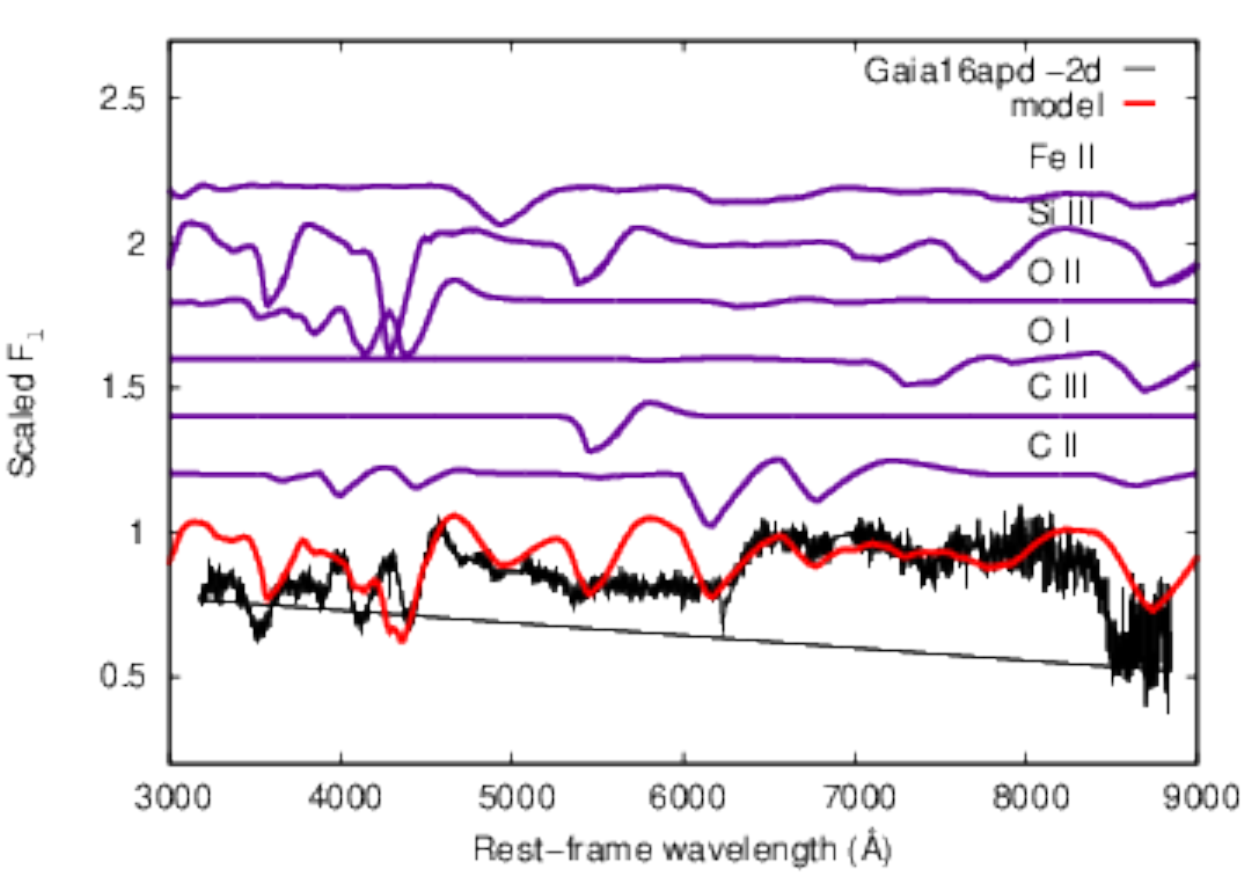}
\includegraphics[width=5cm]{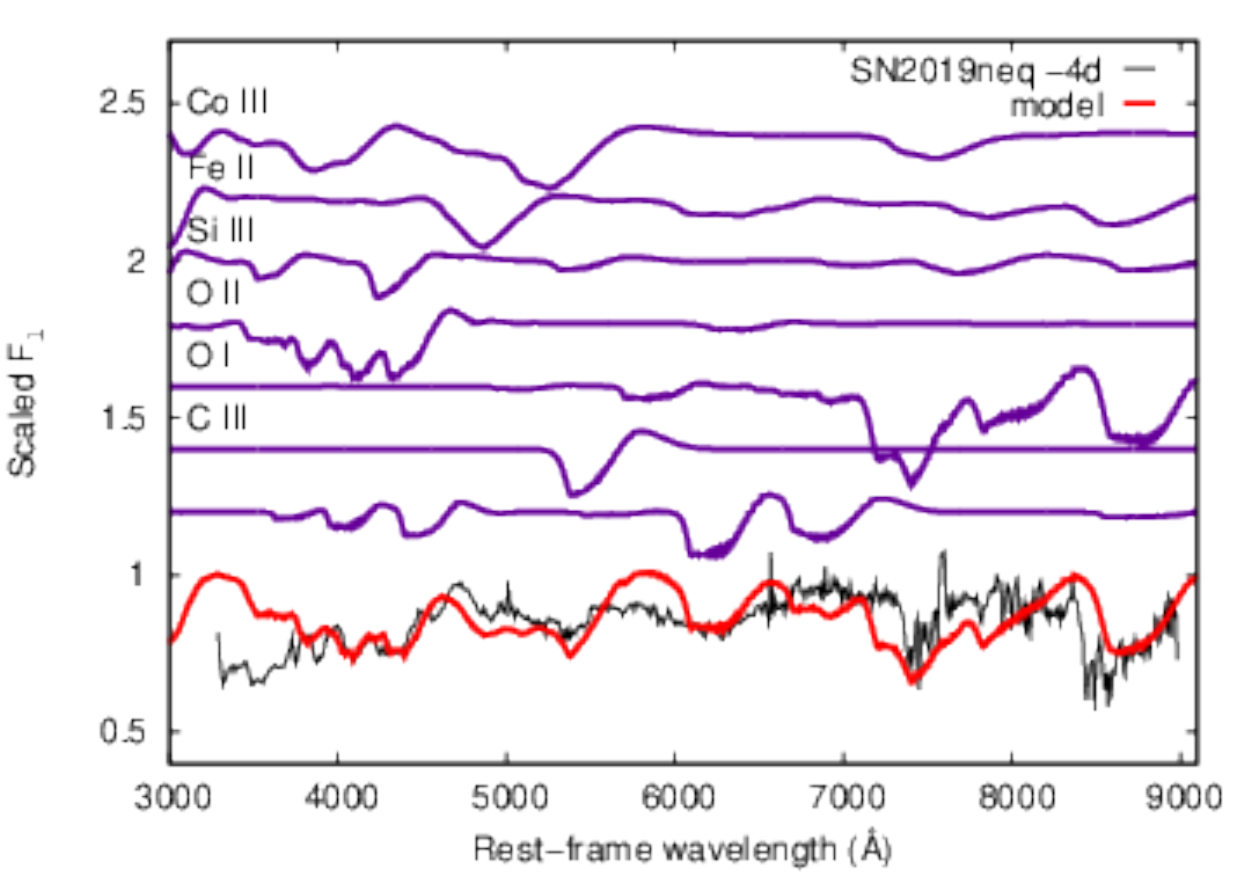}
\includegraphics[width=5cm]{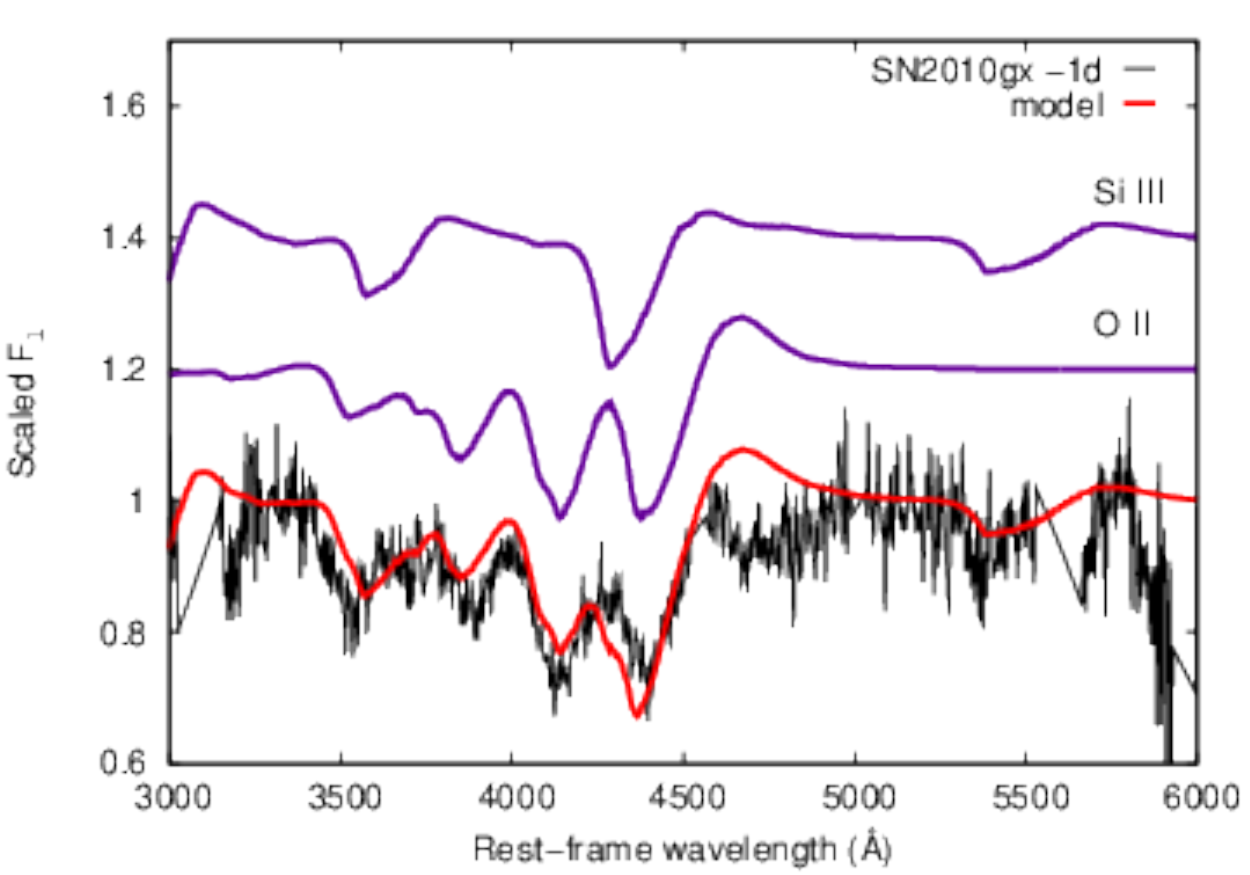}
\includegraphics[width=5cm]{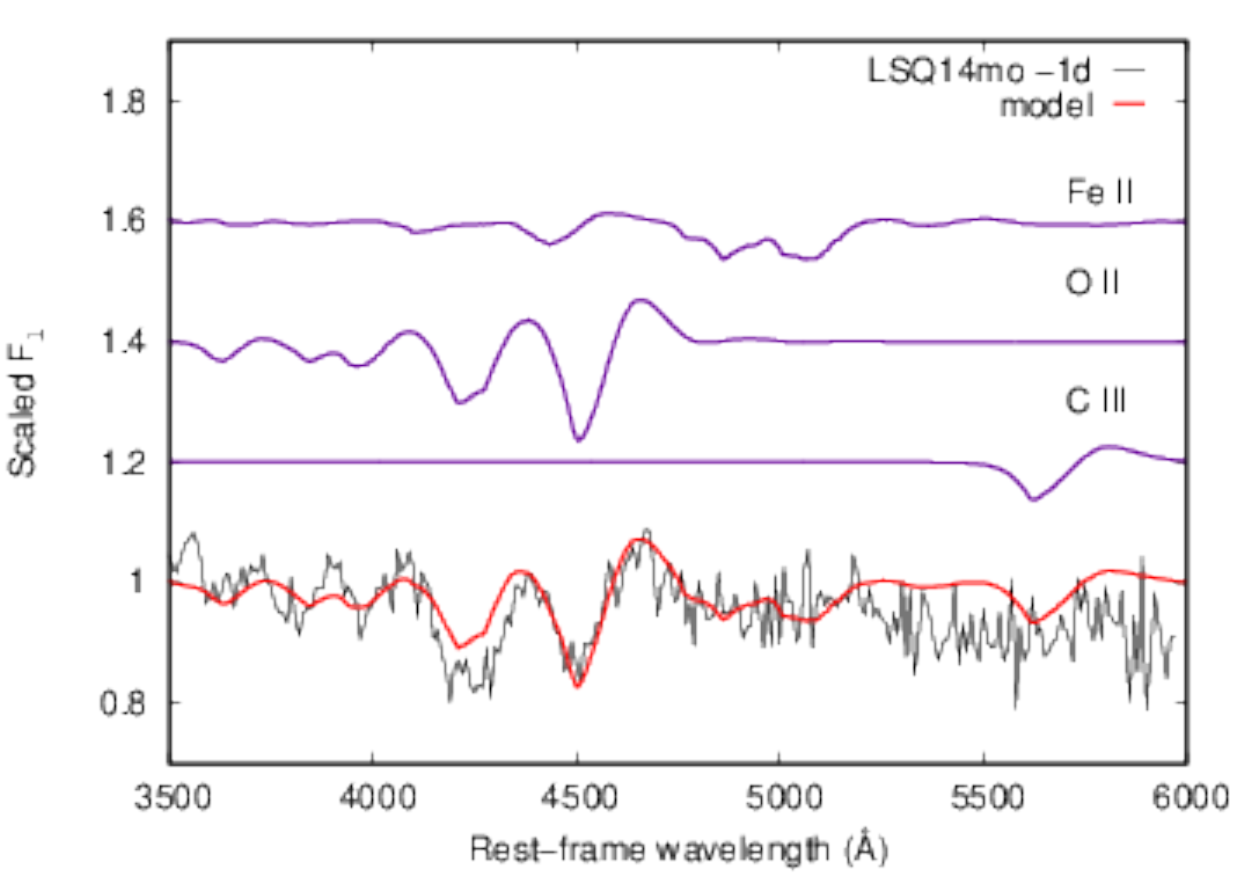}
\includegraphics[width=5cm]{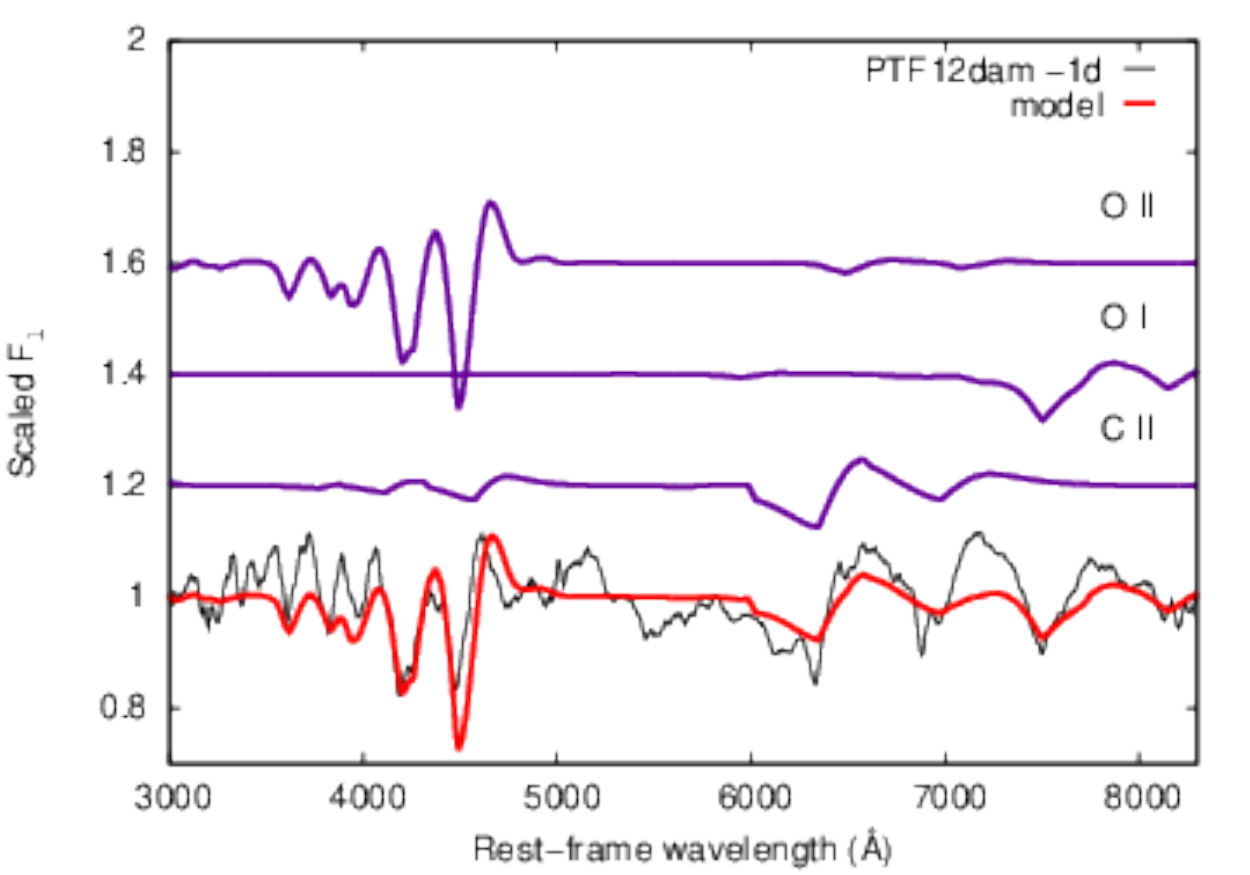}
\caption{Near-maximum spectra of Type W  SLSNe-I  with the same color-coding as Figure \ref{fig:early_wmodels}. }
\label{fig:near_wmodels}
\end{figure*}

Figure \ref{fig:early_wmodels} displays the spectrum modeling of Type W objects that have available data in the early pre-maximum phase up to -7 rest-frame days relative to the moment of the maximum, while  Figure \ref{fig:near_wmodels} shows the near-maximum (from -7 to 0 days from maximum) spectra of the SLSNe-I belonging to the Type W group. The color coding is the same in these figures: black lines denote to the observed continuum-normalized spectra, while red lines are coding the best-fit models obtained in SYN++. The single-ion contributions to the overall model spectra are plotted with purple color, shifted vertically from each other for clarification. 

It is seen that the spectra displayed in Figures \ref{fig:early_wmodels} and \ref{fig:near_wmodels} are dominated by the W-shaped absorption feature at $\sim$3500 - 5000 \AA, which were identified as the lines of O II. In some previous studies, it was shown that this wavelength region can either be fitted with the lines of C III, O III, and Si III though \citep[e.g.][]{ktr20-2}. In addition to the strong O II features, the models also contain C II, C III, Si II, Si III, Fe II, or Co II lines, which are typically present in Type I SLSNe
\citep[e.g.][]{dessart12,mazzali16,dessart19,galyam19b,inserra13}.

Table \ref{tab:osszefoglalo} shows the $v_{\rm phot}$ and $T_{\rm phot}$ values of the best-fit models. The optical depth, line-forming velocity region, $aux$ and $T_{\rm exc}$ parameters for each identified ion are collected in the Table \ref{tab:local_w} of Appendix in case of all Type W  SLSNe-I . 

In is seen in Table \ref{tab:osszefoglalo} that the $v_{\rm phot}$ values of Type W  SLSNe-I  span between a wide range of 12000 and 24000 km s$^{-1}$. According to \citet{inserra18}, the photospheric velocity gradient of SLSNe-I correlates with the $v_{\rm phot}$ near the moment of the maximum. It was confirmed later by e.g. \citet{ktr21}. This means that one spectrum taken before or near maximum is sufficient to decide whether the object has a high or low photospheric velocity gradient, and one may classify it into the spectroscopically Fast (high velocity gradient) or the spectroscopically Slow (low velocity gradient) group created by \citet{inserra18} without owning +30 days phase post-maximum data (see Figure 11 of \citet{ktr21} for further discussion).

 As was mentioned earlier, in the study of \citet{ktr21}, SLSNe-I showing a larger  $v_{\rm phot}$ than 20000 km s$^{-1}$ near maximum were classified as spectroscopically Fast  SLSNe-I , while the objects having $v_{\rm phot} <$  16000 km s$^{-1}$ were called to be spectroscopically Slow evolving. The  SLSNe-I  that were found to be consistent with $v_{\rm phot}~=~$18000 - 20000 km s$^{-1}$, were classified in neither group and were called spectroscopically uncertain (N). As the sample in the present study is the same as in \citet{ktr21}, the classes of each SLSNe-I were adopted from there.  Table \ref{tab:osszefoglalo} shows that Type W  SLSNe-I  are represent themselves in both the high- and the low-velocity gradient group, so they are present in the spectroscopically Slow and Fast evolving classes as well.

Not only velocity gradients but the rise-times of the light curves are used to classify the SLSNe-I into the Fast/Slow evolving subgroup as well. \citet{inserra18} found that the Fast evolving objects rise averagely from the explosion to the maximum in $\sim$ 30 days, while Slow evolving objects tend to have $t_{\rm rise}$ larger than 50 days. Here, we set the borderline between these groups at $t_{\rm rise}~=~$40 days: the SLSNe-I that show $t_{\rm rise}<$40 days are called photometrically Fast evolving, while the objects with $t_{\rm rise}>$40 days are referred as photometrically Slow evolving. As can be seen in Table \ref{tab:osszefoglalo}, in most cases, the photometric evolution time scale agrees with the spectroscopic one, based on the velocity gradients. However, there are 5 objects in the W-group showing a low velocity gradient together with a short rise-time, which means a Slow spectroscopic evolution together with a Fast photometric evolution. According to this finding, it can be said that the Fast/Slow subclass is not necessarily the same in terms of photometry and spectroscopy.  

The photospheric temperatures of the best-fit models of Type W  SLSNe-I  are presented in Table \ref{tab:osszefoglalo} too, which are ranging between 8000 and 20000 K. According to \citet{hatano99}, the singly ionized oxygen may be excited in a large temperature range, which is consistent with the $T_{\rm phot}$ values of the SYN++ models built to synthesize the spectra of Type W  SLSNe-I .

\subsection{Type 15bn objects}

\begin{figure*}
\centering
\includegraphics[width=5cm]{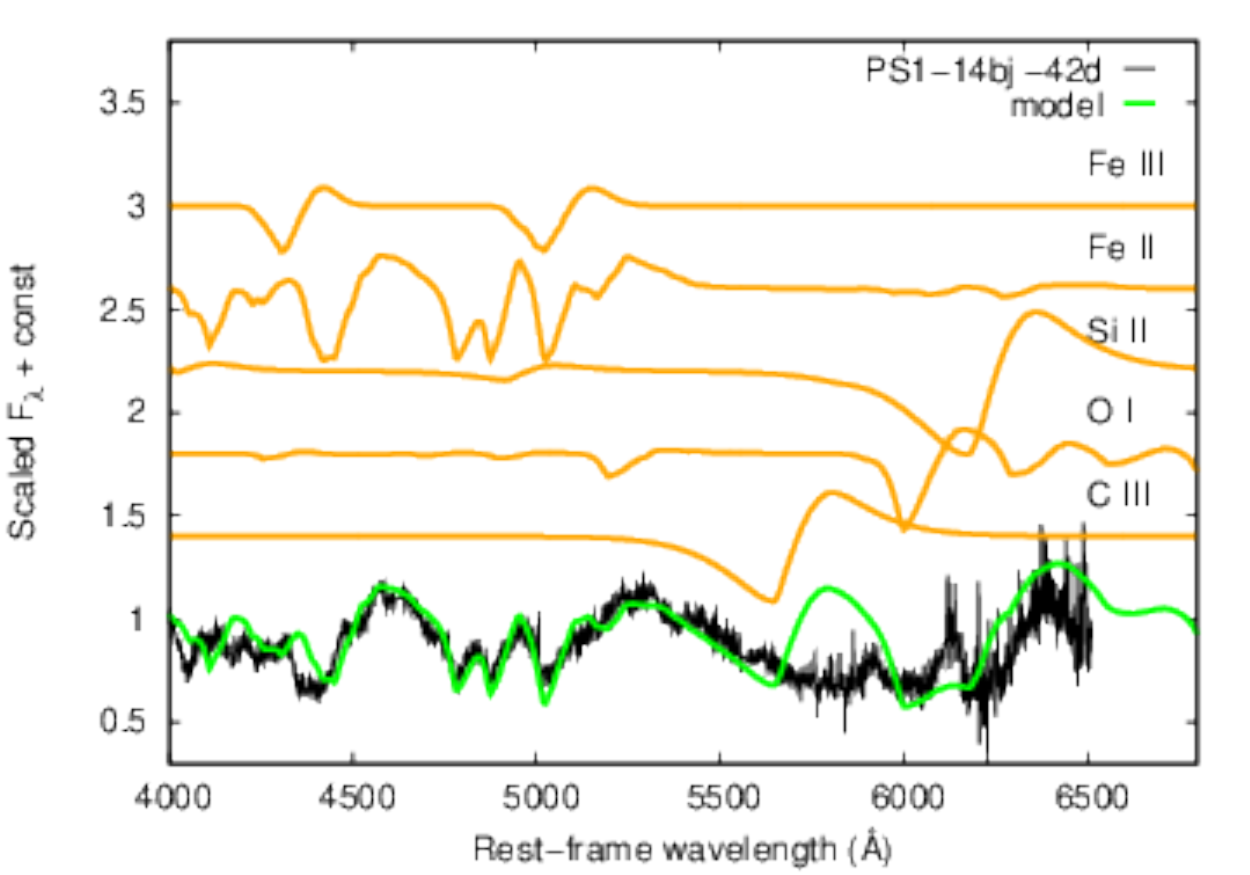}
\includegraphics[width=5cm]{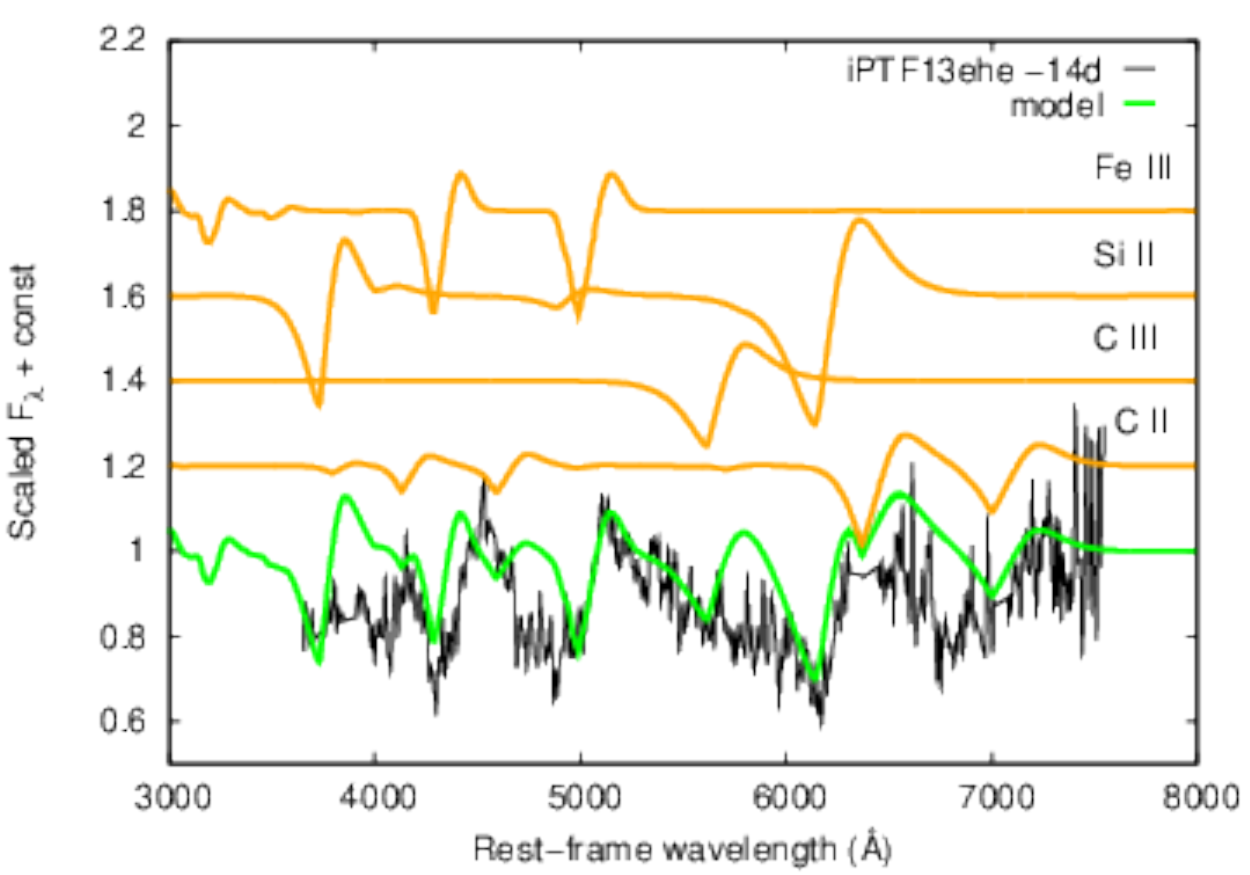}
\includegraphics[width=5cm]{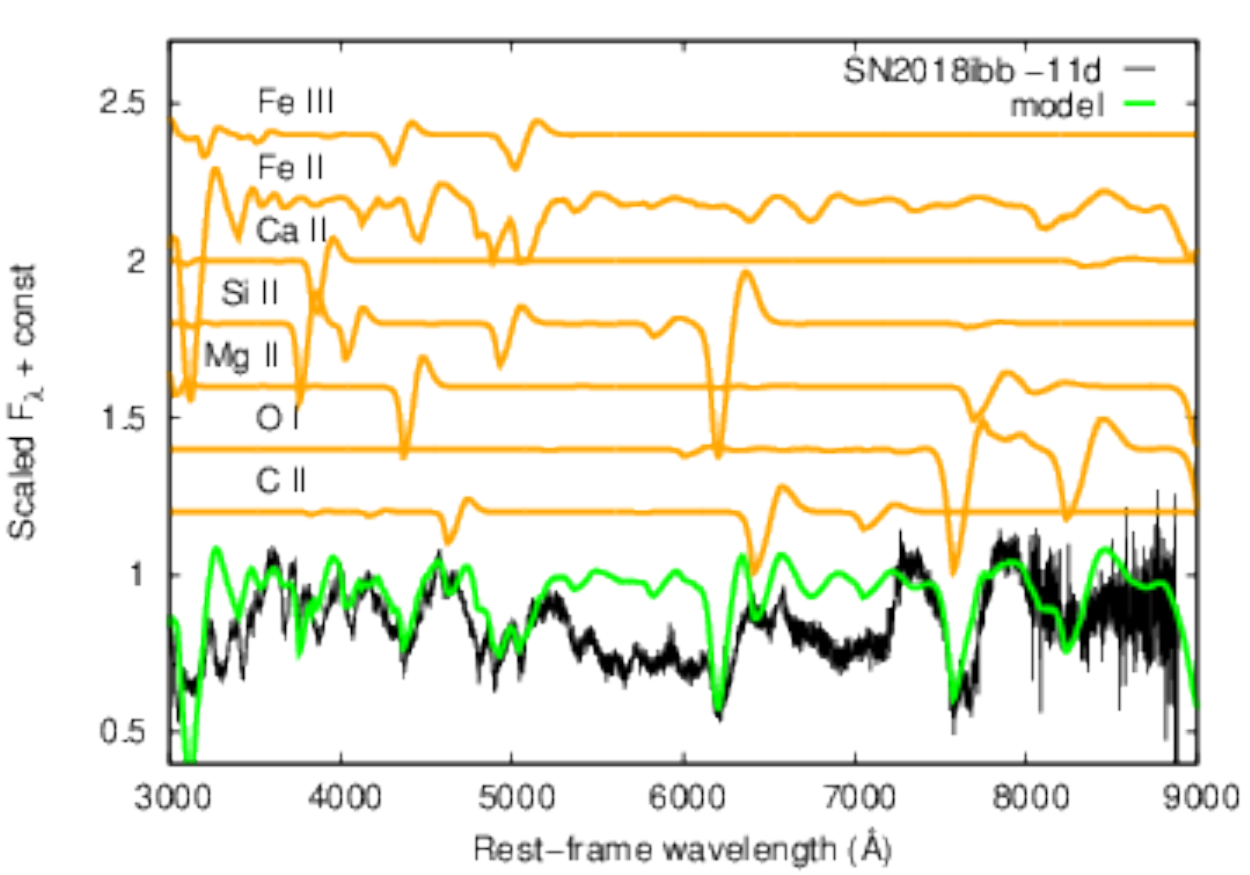}
\caption{Early pre-maximum spectra of Type 15bn  SLSNe-I . The observed spectra (black lines) are plotted together with their best-fit SYN++ model (green lines). The contribution of the identified elements to the overall model spectra are shown with orange color, shifted vertically for clarification. }
\label{fig:early_15bnmodels}
\end{figure*}

\begin{figure*}
\centering
\includegraphics[width=5cm]{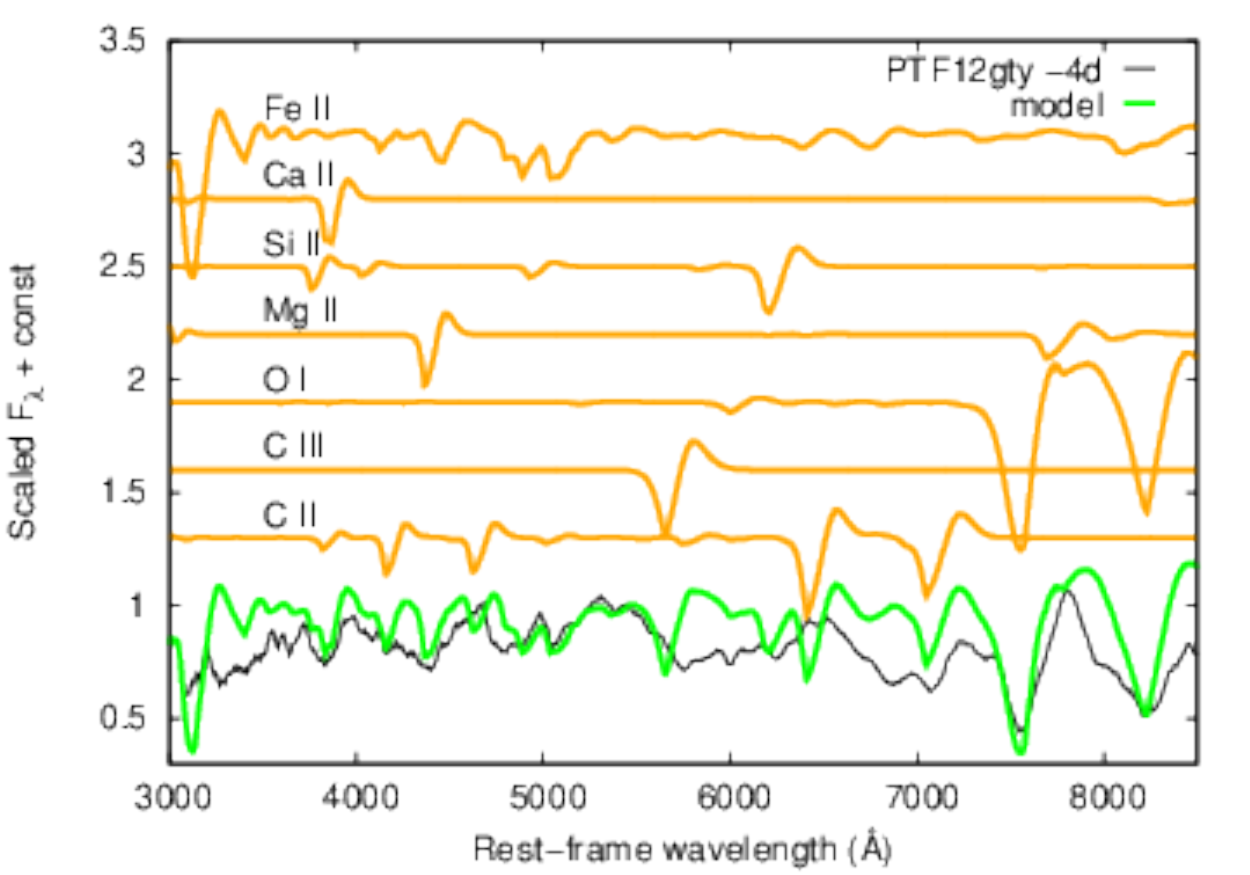}
\includegraphics[width=5cm]{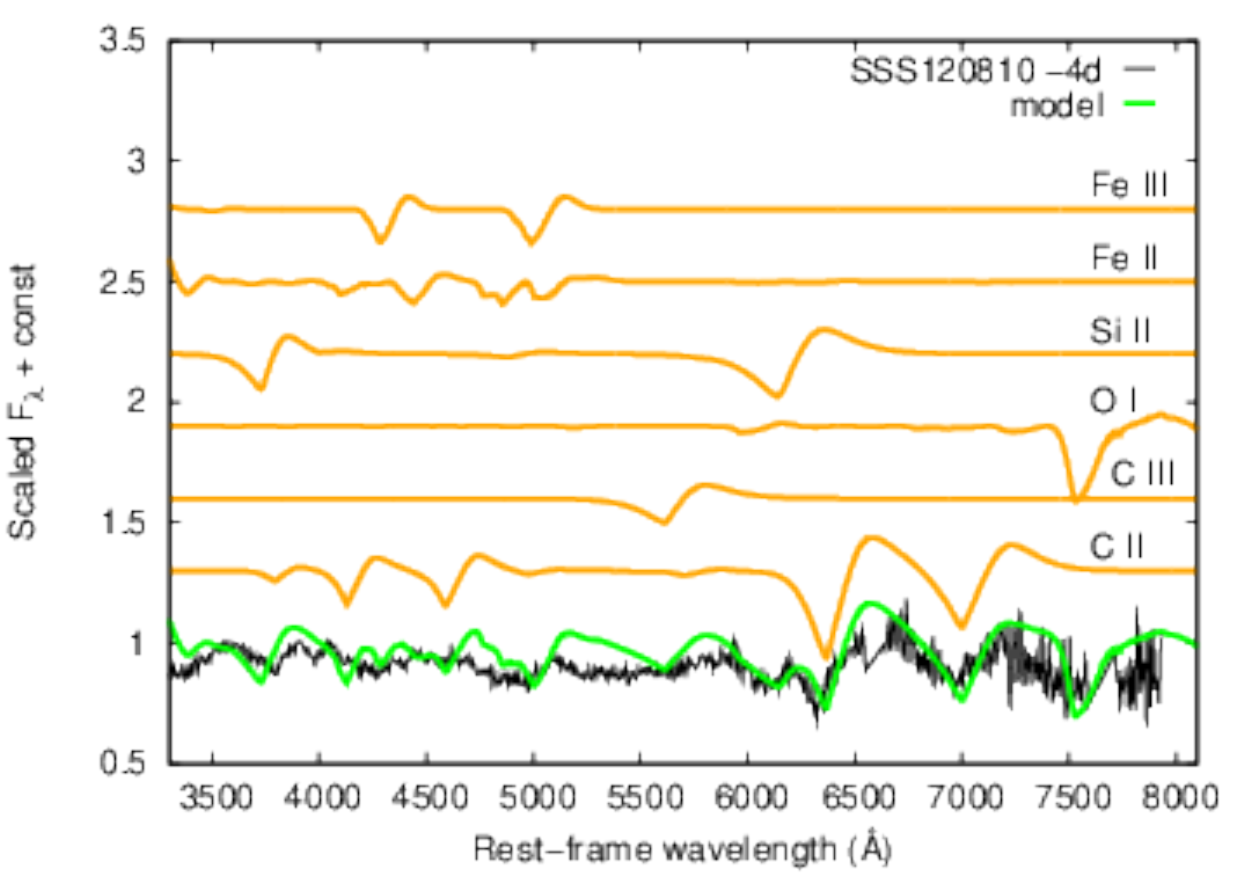}
\includegraphics[width=5cm]{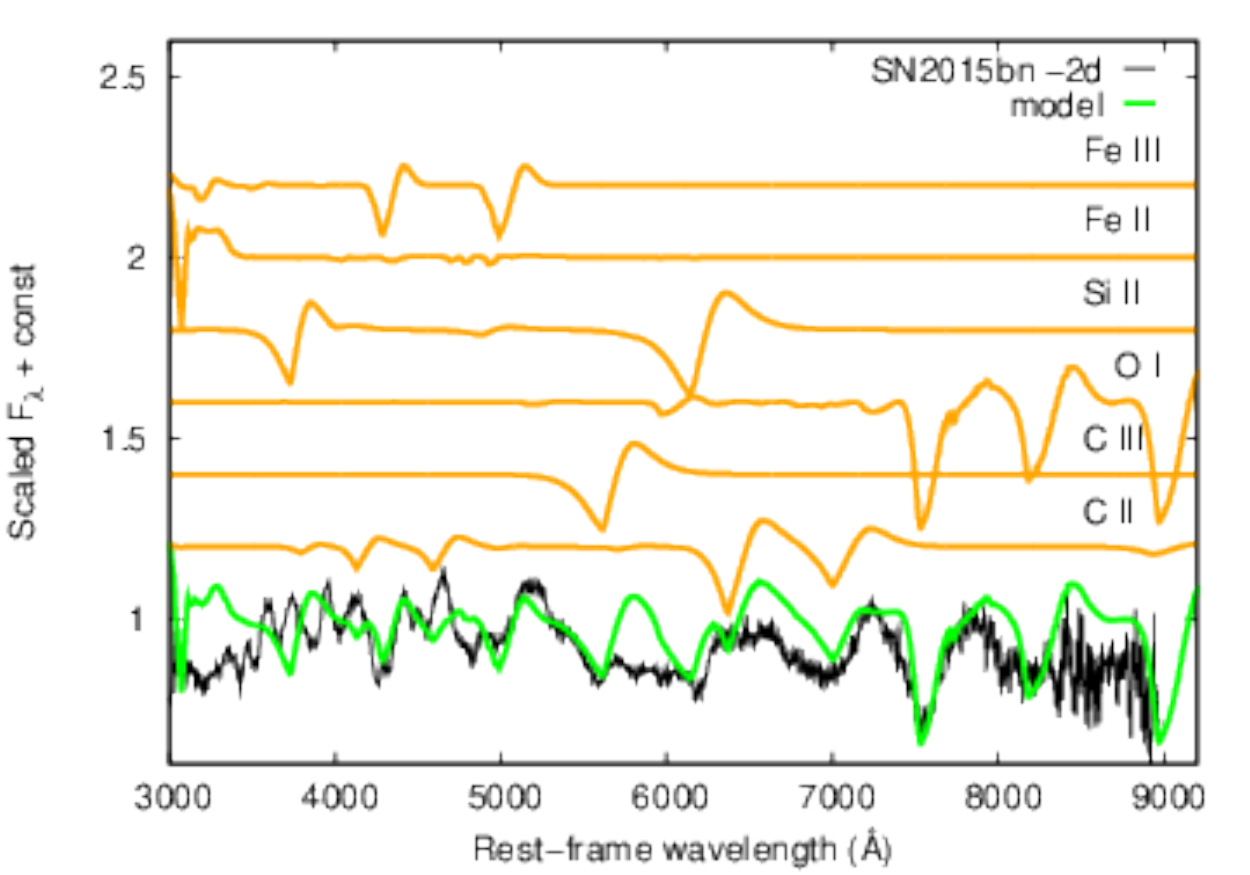}
\includegraphics[width=5cm]{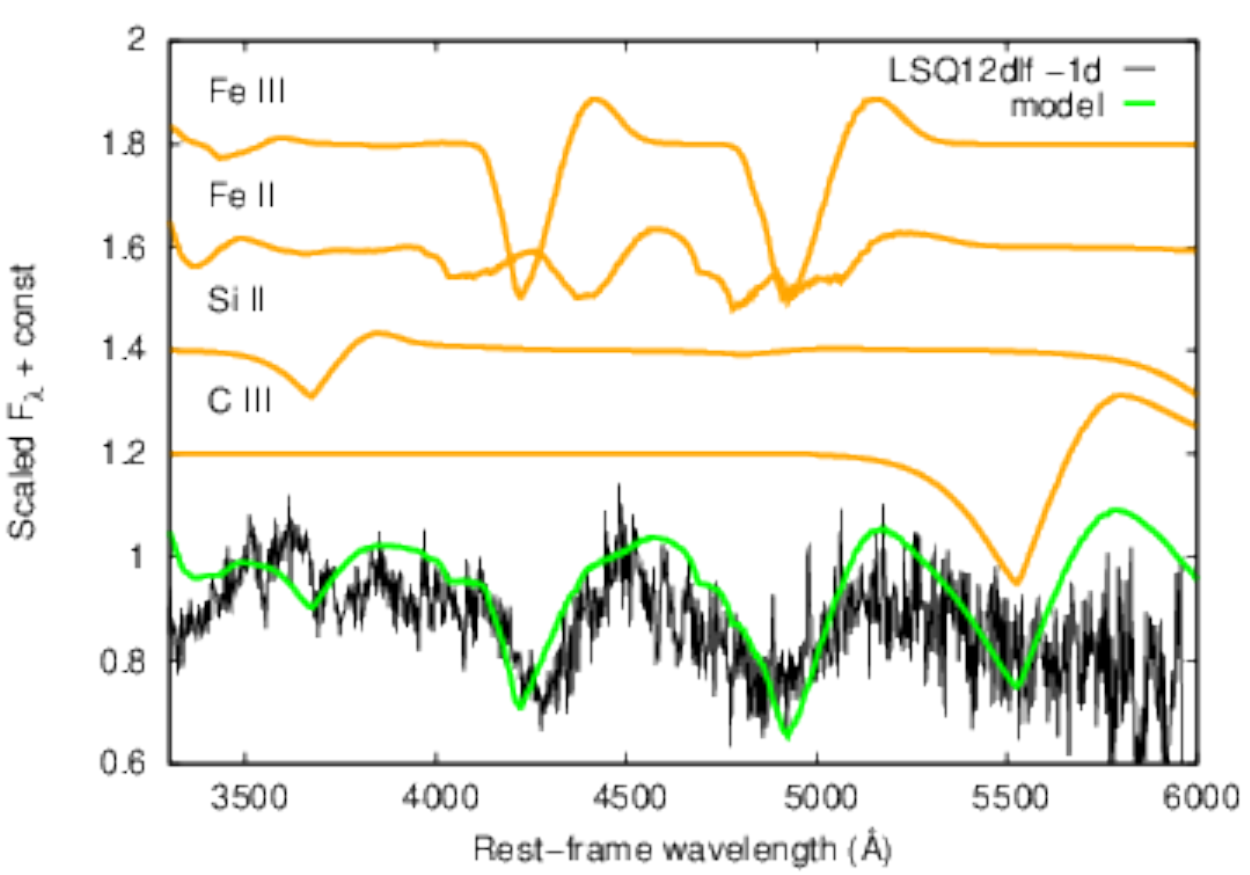}
\includegraphics[width=5cm]{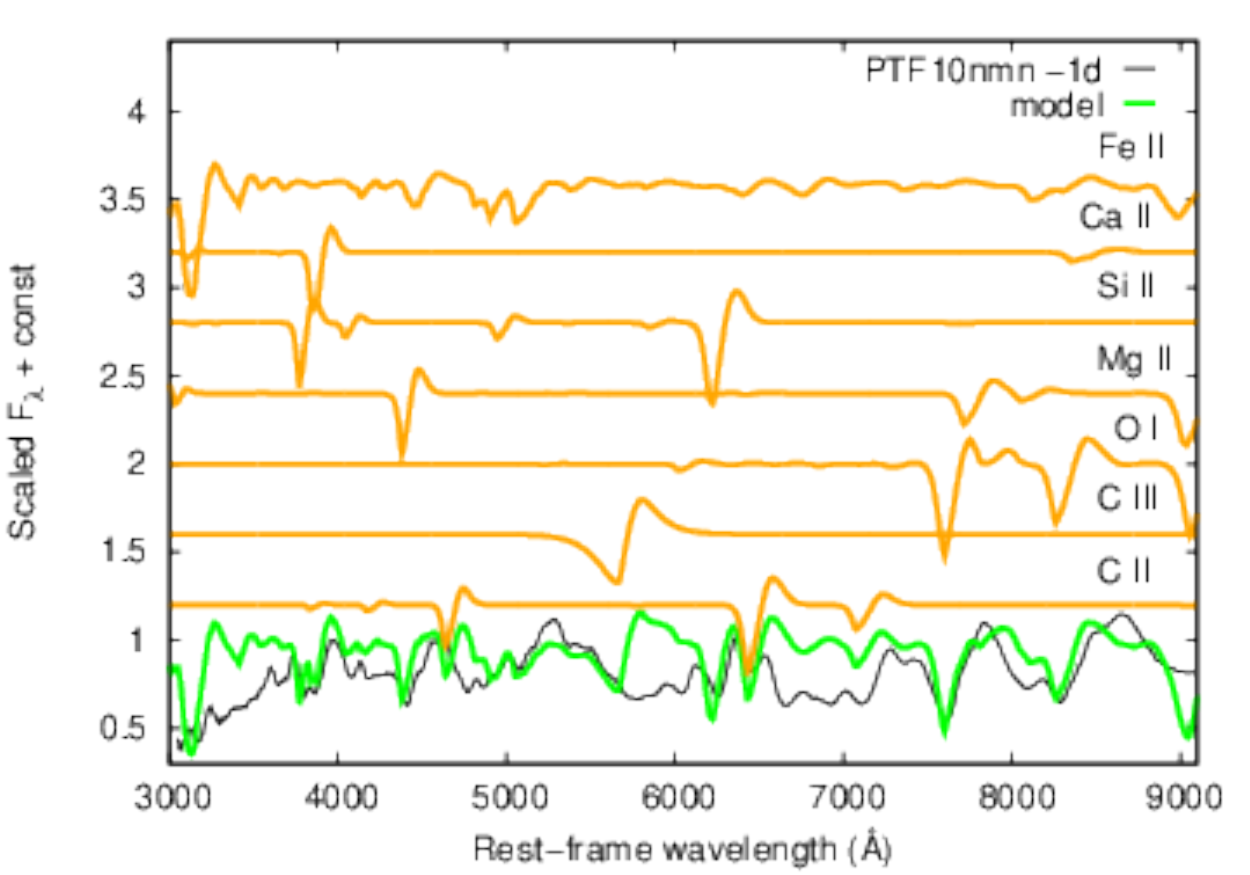}
\includegraphics[width=5cm]{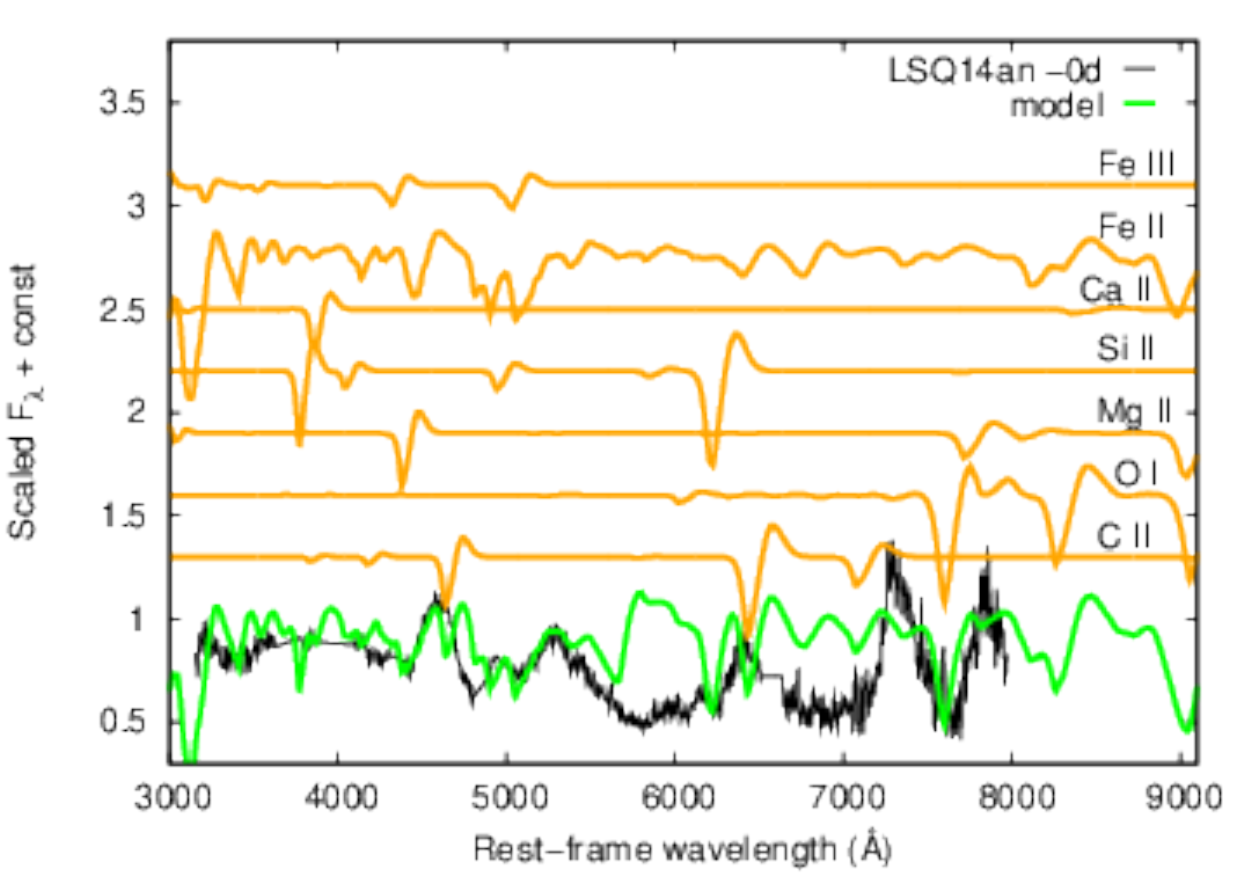}
\caption{Near-maximum spectra of Type 15bn  SLSNe-I . The color coding is the same as in Figure \ref{fig:early_15bnmodels}.}
\label{fig:near_15bnmodels}
\end{figure*}

Pre-maximum spectrum modeling of Type 15bn SLSNe-I was carried out as well using SYN++, which are plotted  for the early pre-maximum and the near maximum phases in Figure \ref{fig:early_15bnmodels} and \ref{fig:near_15bnmodels} respectively. It is seen that the best-fit models do not contain O II lines at all, because of the absence of the W-shaped absorption features. There is some neutral oxygen modeled at the redder wavelengths though, so the element of oxygen is not missing from the spectrum, just from the bluer end of the spectrum. In addition, the best-fit models contain C II, C III, O I, Na I, Mg II, Si III, Ca II, and Fe II lines. 

The best-fit $v_{\rm phot}$ and $T_{\rm phot}$ values obtained for Type 15bn  SLSNe-I  are shown in Table \ref{tab:osszefoglalo}, while ion-specific data of the best-fit SYN++ models are collected in Table \ref{tab:local_15bn} of the Appendix. 

It is seen in Table \ref{tab:osszefoglalo} that the photospheric velocities of the Type 15bn SLSNe-I are ranging from 8000 to 15000 km s$^{-1}$, indicating that all of them belong to the spectroscopically Slow evolving group, as was classified by \citet{ktr21}. This classification is consistent with the photometric evolution in 7 cases, while  in case of 2  SLSNe-I(SSS120810 and LSQ14an)  the rise-time could not be determined precisely. For SSS120810, the early pre-maximum phases weren't observed photometrically \citep[see e.g.][]{nicholl14}, while LSQ14an was discovered after peak \citep[see][]{inserra17}. Therefore the photometric classification of these objects remains uncertain, thus marked with a '?' symbol in Table \ref{tab:osszefoglalo}.

The best-fit $T_{\rm phot}$ values seem to be generally lower compared to Type W  SLSNe-I , ranging between 8000 and 12000  K, so one may say that the main difference of Type W and Type 15bn  SLSNe-I  can be found in the difference in the photospheric temperature. However, according to \citet{hatano99}, this temperature range is consistent with the presence of the O II lines as well, so there has to be other reasons apart from the temperature for the absence of the W-shaped feature from the spectra of Type 15bn objects.

\section{Discussion}\label{sec:disc}

In this section, we discuss some possible explanations for the differences in the pre-maximum spectrum evolution of Type W and Type 15bn  SLSNe-I . 

\subsection{Temperature/ion composition}

As mentioned in Section \ref{subsec:syn}, it is important to note that the photospheric temperatures obtained from SYN++ modeling have to be treated with caution apart from the fact that the relative fluxes downloaded from the Open Supernova Catalog of the examined SLSNe-I were assumed to be calibrated correctly and the spectra were corrected for MW extinction. Most pre-maximum spectra obtained for the SLSNe-I in our sample shows a hot blue continuum (see Figure \ref{fig:classes}) consistently with the behavior of other SLSNe-I. This suggests that in most cases, the extinction of the host galaxy is negligible, and the fitted $T_{\rm phot}$ values are reasonable. However, there are a few objects (e.g. SN~2018hti), where the continuum is redder. In these cases, the host galaxy may show a more significant reddening, which makes the fitted $T_{\rm phot}$ more uncertain. 

\begin{figure*}
\centering
\includegraphics[width=15cm]{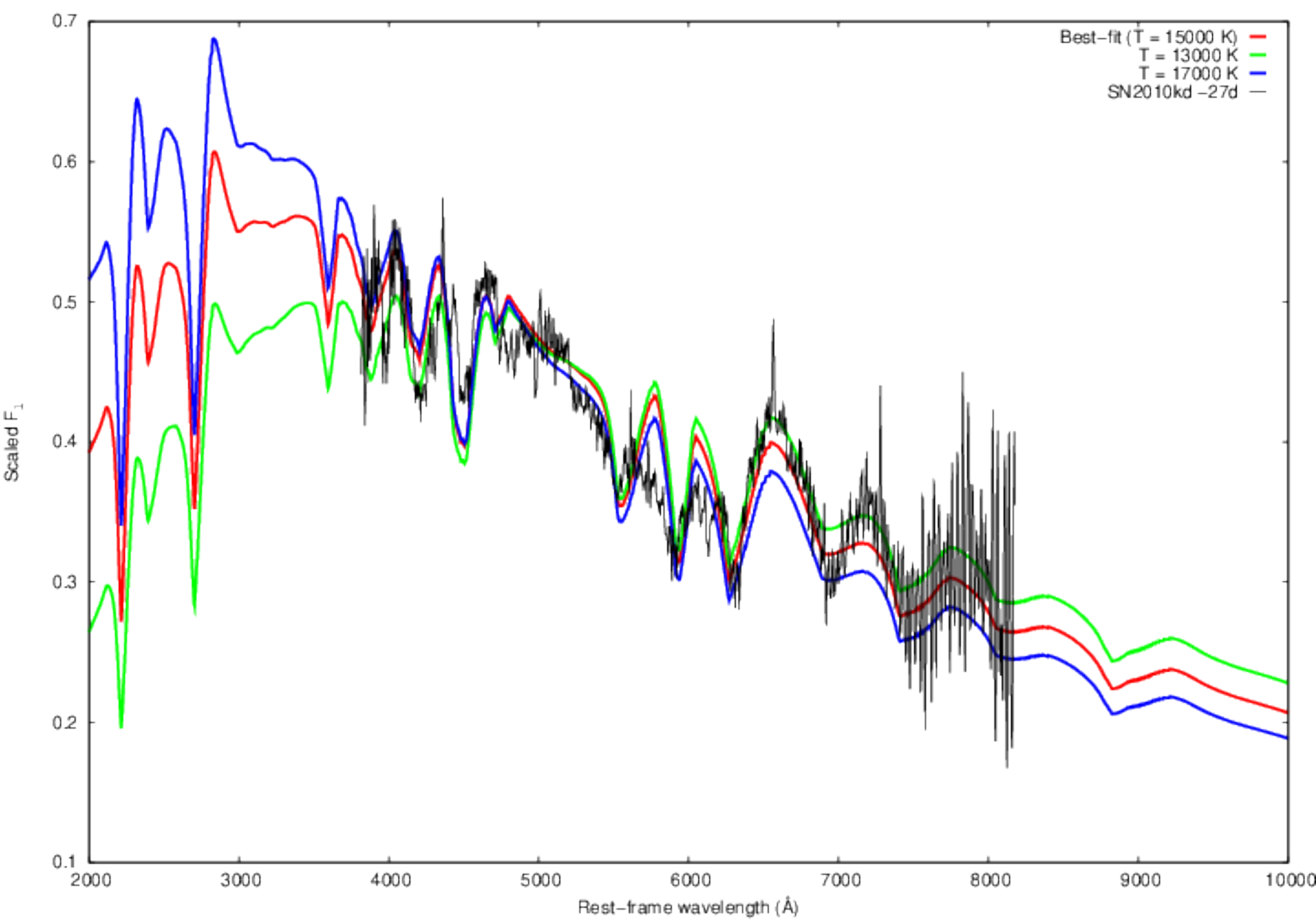}
\caption{The -22 days phase observed spectrum of SN~2010kd (black) together with its best-fit SYN++ model showing  $T_{\rm phot}~=~$15~000 K (red). A lower ($T_{\rm phot}~=~$13~000 K), and a higher ($T_{\rm phot}~=~$17~000 K) model is plotted as well with green and blue colors respectively. The three models share the same global and local parameters with the exception of the photospheric temperature.}
\label{fig:tph}
\end{figure*}

On Figure \ref{fig:tph}, the -22 days phase spectrum of SN~2010kd is displayed together with its best fit model having  $T_{\rm phot}~=~$15~000 K. Two additional models having the same parameter values with the exception of $T_{\rm phot}$ are plotted with different colors. Green color codes the $T_{\rm phot}~=~$13~000 K model, while the blue line denotes to the $T_{\rm phot}~=~$17~000 K model. It is seen that the lower temperature model overestimates the flux on the redder part of the observed spectrum and underestimates the fluxes at the bluer wavelengths. The behavior of the higher temperature model is the opposite. From this fact we conclude that the uncertainty of the temperature at the photosphere synthesized using SYN++ has an uncertainty of $\sim$ 2000 K.

\begin{table*}
\tiny
\caption{Photospheric velocities and temperatures of the studied Type W and Type 15bn  SLSNe-I  sorted by the pre-maximum phase. Classification into spectroscopically Fast (high velocity gradient) and spectroscopically Slow (low velocity gradient) is shown as well, compared to the photometric Fastness/Slowness divided by the rise-time of the light curve. The 8th column shows if the photometric and the spectroscopic class of a specific agrees (+ symbol) or not (- symbol). The 9th column shows if the given SLSN-I has some results from polarimetric measurements showing a null-polarization, or polarization increasing with time, which accounts for the geometry of the explosion.  In the 10th column, the "bumpiness" of the pre-maximum LC is shown with + and - symbols. The photospheric velocity gradients measured between the moment of the maximum and $\sim$ 30 days after are shown as well in case of the 9 objects in the sample that had available +30 days post-maximum spectroscopic data in the OSC.  }
\label{tab:osszefoglalo}
\centering
\begin{tabular}{lcccccccccc}
\hline
\hline
SN & Phase (1) & $v_{\rm phot}$ (2) & $T_{\rm phot}$ (2)& $t_{\rm rise}$ (1) & Sp. type (1) & Ph. type (1) & Agree (2) & Polarization &  Early bump &  v$_{\rm grad}$ (1)\\
   & days & km s$^{-1}$ & K & days &  F/S/N  & F / S & +/- & Null/Increasing  &  +/ - &  km s$^{-1}$ day$^{-1}$  \\
\hline
\multicolumn{11}{c}{Type W} \\
\hline
SN~2018hti & -54 & 15000 & 13000 & 97 & S & S & + & Null (3) &  + (9, 10) & --\\
DES14X3taz & -29 & 13000 & 15000 & 45 & S & S & + & -- & + (11) & --\\
SN~2010kd & -22 & 11000 & 15000 & 48 & S & S & + & -- &  - & --\\
PTF09atu & -19 & 10000 & 12000 & 42 & S & S & + & -- &  - & -- \\
SN~2018bsz & -16 & 10000 & 12000 & 76 & S & S & + & Increasing? (4) &  + (12, 13, 14) & -- \\
PTF09cnd & -14 & 13000 & 14000 & 54 & S & S & + & -- &  - &  133\\
LSQ14bdq & -11 & 11000 & 14000 & 47 & S & S & + & -- &  + (15) & -- \\
SN~2011kg & -10 & 11500 & 12000 & 26 & S & F & - & --&  -& -- \\
SN~2006oz & -5 & 15000 & 15000 & 26 & S & F & - & -- &  + (16)& -- \\
iPTF13ajg & -5 & 10000 & 12000 & 47 & S & S   & + & --&  -& -- \\
SN~2016els & -5 & 19000 & 20000 & 22 & N & F  & - & --&  -& -- \\
SN~2019neq & -4 & 24000 & 14000 & 28 & F & F  & + & --&  - &  400\\
SN~2016ard & -4 & 14000 & 20000 & 25 & S & F  & - & --&  - &  80 \\
SN~2005ap & -3 & 23000 & 20000 & 19 & F & F  & + & --&  -\\
 Gaia16apd & -2 & 20000 & 20000 & 19 & F  & F  & + & --& - &  330 \\
SN~2010gx & -1 & 20000 & 17000 & 25 & F & F  & + & --&  - &  310 \\
LSQ14mo & -1 & 10000 & 12000 & 27 & S &  F  & - & Null (5)&  - & -- \\
PTF12dam & -1 & 12000 & 14000 & 63 & S & S  & + & Null (6) &  + (17)& -- \\
\hline
\multicolumn{11}{c}{Type 15bn} \\ 
\hline
PS1-114bj & -42 & 10000 & 10000 & 139 & S & S & + & --&  -& -- \\
SN~2015bn & -17 & 10000 & 12000 & 91 & S & S & + & Increasing (7, 8)&  - &  40 \\
iPTF13ehe & -14 & 10000 & 11000 & 83 & S & S & + & --&  - & -- \\
SN~2018ibb & -11 & 8000 & 11000 &  112 & S & S & + & -- &  -& -- \\
PTF12gty & -4 & 8000 & 8500 & 49 & S & S & + & -- &  -& -- \\
SSS120810 & -1 & 10000 & 10000 & 32  ?  & S &  ? &  ? & --&  - &  50\\
LSQ12dlf & -1 & 15000 & 11000 & 42 & S & S & + & --&  - &  200\\
PTF10nmn & -1 & 8000 & 8000 & 105 & S & S & + & --&  - &  120\\
LSQ14an & -0 & 8000 & 8000 & 18  ? & S &  ? &  ? & -- &  - & -- \\
\hline
\end{tabular}
\tablecomments{
1: Adopted from \citet{ktr21}; 
2: Present paper; 
3: \citet{lee19};
4: \citet{maund21};
5: \citet{lelo15b};
6: \citet{cikota18};
7: \citet{inserra16};
8: \citet{lelo17};
 9: \citet{lin20};
10: \citet{fiore22};
11: \citet{smith16};
12: \citet{anderson18};
13: \citet{chen21};
14: \citet{pursi22};
15: \citet{nicholl15};
16: \citet{lelo12};
17: \citet{vrees17}
}
\end{table*}

\begin{figure*}
\centering
\includegraphics[width=10cm]{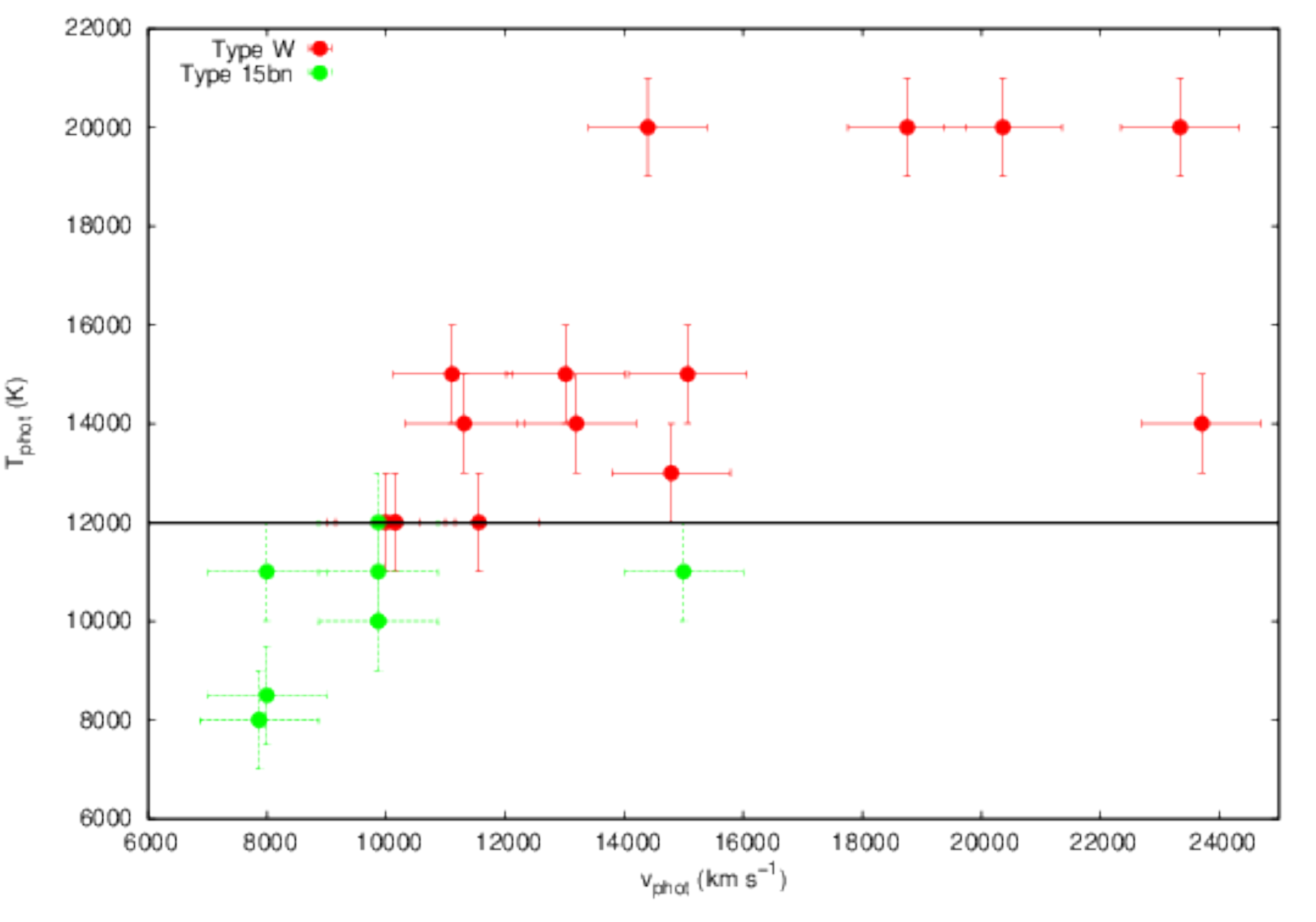}
\caption{Photospheric velocities as a function of photospheric temperatures of the 27 studied  SLSNe-I  obtained from SYN++ modeling. Type W SLSNe-I are plotted with red, while Type 15bn objects are shown with green dots.}
\label{fig:sebessegek}
\end{figure*}

It was found in Section \ref{sec:model} that Type W SLSNe-I globally show larger photospheric velocities in the pre-maximum phase compared to Type 15bn objects.
 Therefore one may use the value of $v_{\rm phot}$ to distinguish between these two sub-classes. Another global parameter of the SYN++ code that can help to separate Type W and Type 15bn SLSNe-I is the photospheric temperature.
Not only the fitted blackbody continuum, but also the ion composition of each spectrum can be a tracer of the real photospheric temperature, since, in case of thermal processes, the higher the excitation/ionization of a specific element, the higher the temperature at the photosphere. This is consistent with the finding that the globally hotter Type W SLSNe-I show singly ionized oxygen lines, while only neutral oxygen features seem to present in the spectra of the globally cooler Type 15bn objects. 

Figure \ref{fig:sebessegek} displays the photospheric velocities obtained from the SYN++ modeling as a function of the photospheric temperatures. Type W SLSNe-I are plotted with red, while green dots denote to the Type 15bn SLSNe-I. It is seen  Type 15bn  SLSNe-I , which show lower velocities before maximum, indeed tend to have lower temperatures ($T_{\rm phot} < $12000 K) compared to Type W  SLSNe-I  ($T_{\rm phot} >$ 12000 K). 

\begin{figure*}
\centering
\includegraphics[width=20cm]{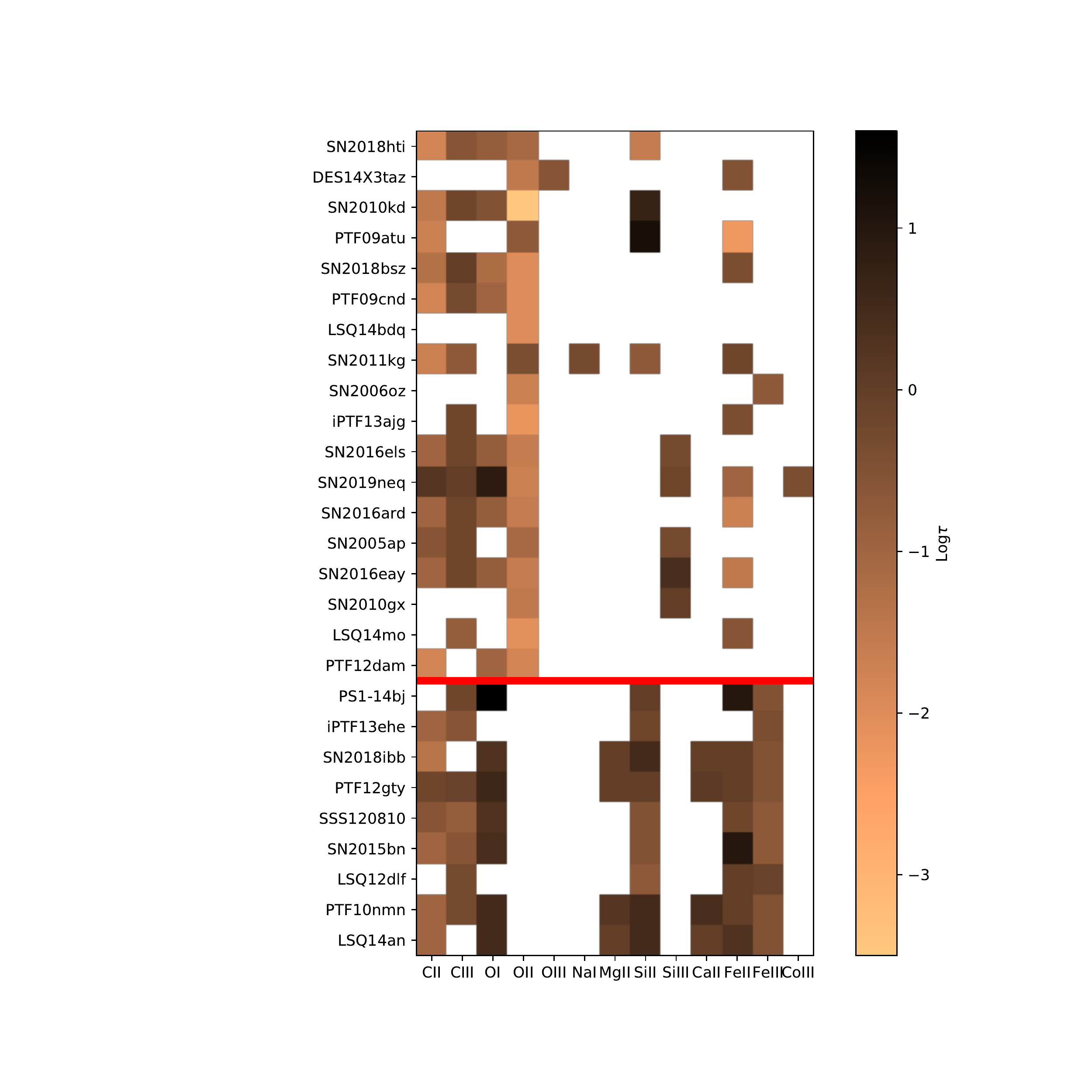}
\caption{Chemical composition of the examined 27 SLSNe-I. On the vertical axis, the names of the studied SLSNe-I are listed, while the horizontal axis lists the possible identified ions.  The horizontal red line divides Type W (above) and Type 15bn (below) SLSNe-I. }
\label{fig:ionosszetetel}
\end{figure*}

The chemical composition of the objects is in connection with the photospheric temperatures, therefore we examined the sample in this aspect also. Figure \ref{fig:ionosszetetel} displays the identified ions in the spectra on the horizontal axis, while on the vertical axis, the names of the individual objects are plotted. The colors of the figure allude to the optical depth of an ion: the darker the color, the larger the optical depth. This way one can easily see the strength of the ions identified in the spectra. Above the red horizontal dividing line, Type W  SLSNe-I  are plotted, while below this line, Type 15bn objects are shown.

In the followings, we discuss the effect of each individual ion in the spectrum formation of Type W and Type 15bn  SLSNe-I . 

\begin{itemize}
    \item {\bf Carbon:} 
    \citet{galyam19} showed that in addition to the typically present oxygen features, SLSNe-I tend to show carbon lines before and around the peak luminosity. It was suggested that the presence of these dominant C-lines is caused by the emission of a mostly pure C/O envelope that does not contain a significant amount of higher elements mixed up from the deeper layers of the ejecta. This is consistent with the strong presence of C II and C III lines in the spectra of the examined SLSNe-I, belonging to either the Type W or the Type 15bn group. 
    \item {\bf Oxygen:} The spectra of almost all SLSNe-I is usually dominated by strong oxygen features \citep[e.g.][]{quimby11,mazzali16,liu17,quimby18}. It is present in the pre-maximum of 26 examined SLSNe-I out of 27. It can be seen in Figure \ref{fig:ionosszetetel} that Type W objects show O I, O II, and in the case of DES14X3taz, O III lines together, however, the O II lines are missing from the spectra of Type 15bn SLSNe-I in spite of the strong presence of O I. It was found earlier by \citet{mazzali16}
    that the O II lines near 4000-5000 \AA\ are due to non-thermal excitation and can be characterized with a temperature between 12000 and 15000 K \citep{inserra19}. In accordance with this finding, the $T_{\rm phot}$ values of Type 15bn  SLSNe-I  were found to be smaller or equal to 12000K, while Type W objects were modeled applying larger temperatures, in some cases even exceeding 15000 K. 
    \citet{gutierrez22} studied the specific Ic-like SLSN-I, SN~2020wnt in their recent paper, and found that this object does not show O II lines in its pre-maximum spectrum. The $T_{\rm phot}$ of the object was modeled to be 10000 K, which is consistent with the conclusions of \citet{inserra19} and the present paper.
    In our sample, 7 Type W SLSNe-I were modeled with a $T_{\rm phot}$ of 12000 K: PFT09atu ($T_{\rm phot}~=~$12000K), SN~2018bsz ($T_{\rm phot}~=~$12000K), SN2011kg ($T_{\rm phot}~=~$12000K), SN2013ajg ($T_{\rm phot}~=~$12000K), LSQ14mo ($T_{\rm phot}~=~$12000K).
    Thus we conclude that 12000 K is the temperature border between Type W and Type 15bn  SLSNe-I . However, the sample studied here is still constrained to a few objects, thus the examination of a larger sample is needed to strengthen this statement. 
    \item{\bf Sodium:} This element was identified only in case of the Type W SN~2011kg ($T_{\rm phot}~=~$12000K)  to fit the strong absorption feature at $\sim$8000\AA. However, putting Na I into the models entrails the presence of  the Na I D line, which is a resonance line that cannot be made weaker by changing the photospheric temperature.
    \item{\bf Magnesium:} The features of Mg II were identified in case of four Type 15bn objects, but no Type W SLSNe-I: SN~2018ibb ($T_{\rm phot}~=~$11000K), PTF12gty ($T_{\rm phot}~=~$8500K), PTF10nmn ($T_{\rm phot}~=~$8000K) and LSQ14an ($T_{\rm phot}~=~$8000K). According to \citet{hatano99}, the lines of Mg II are the strongest at $\sim$8000 K, which is consistent with our modeling. Thus the presence of Mg II may be a difference between Type W and Type 15bn  SLSNe-I  as well. 
    \item{\bf Silicon:} In accordance with the photospheric temperature, Si II features are present in case of all Type 15bn  SLSNe-I  and a few Type W objects that show relatively low photospheric temperature: SN~2018hti: ($T_{\rm phot}~=~$13000K), SN~2010kd ($T_{\rm phot}~=~$15000K), PTF09atu ($T_{\rm phot}~=~$12000K), SN~2011kg ($T_{\rm phot}~=~$11000K). The Type W objects that were modeled utilizing a higher photospheric velocity value, show Si III instead of Si II:  SN~2016els ($T_{\rm phot}~=~$20000K), SN~2019neq ($T_{\rm phot}~=~$14000K), SN~2005ap ($T_{\rm phot}~=~$20000K),  Gaia16apd ($T_{\rm phot}~=~$20000K) and  SN~2010gx ($T_{\rm phot}~=~$17000K). According to our modeling, the silicon lines are playing a major role in the spectrum formation of SLSNe-I, and support the expectation that Type W  SLSNe-I  are globally showing a larger photospheric velocity compared to Type 15bn  SLSNe-I .
    \item{\bf Calcium:} The Ca II lines are missing from all models of Type W  SLSNe-I , however, they are present in case of 4 Type 15bn objects that had Mg II lines identified in their spectra: SN~2018ibb ($T_{\rm phot}~=~$11000K), PTF12gty ($T_{\rm phot}~=~$8500K), PTF10nmn ($T_{\rm phot}~=~$8000K) and LSQ14an ($T_{\rm phot}~=~$8000K). The presence/absence of Ca II lines from the observed spectra may be affected by the phenomena described in Section \ref{subsec:syn}, thus we cannot conclude that Ca II can be used to distinguish between Type W and Type 15bn SLSNe-I. 
    \item{\bf Iron:} According to \citet{hatano99} the lines of Fe II and Fe III have nearly equal strength in optical depth at $T_{\rm phot}\sim$10000 K, therefore the presence of these iron lines is consistent with either Type W or Type 15bn  SLSNe-I . While Type W  SLSNe-I  mostly show Fe II features, Fe II and Fe III were modeled together in all Type 15bn SLSNe-I with the exception of iPTF13ehe ($T_{\rm phot}~=~$11000K) modeled with only Fe III. 
    \item{\bf Cobalt}: The features of Co III were only modeled in case of SN~2019neq ($T_{\rm phot}~=~$14000K), which is in accordance with \citet{hatano99} suggesting that Co III shows the strongest lines near $T_{\rm phot}~=~$14000K.  
\end{itemize}

\subsection{Pre-maximum light curve undulations}

The main goal of this study is to investigate the dichotomy between the Type W and Type 15bn SLSNe-I subtypes, thus, taking into account the pre-maximum photometric evolution may be an important aspect to explain the spectroscopic bimodality. 

It was shown previously in the literature that some SLSNe-I show undulations in their pre-maximum LCs, often referred as "early bumps". On the contrary, normal Type Ic SNe tend to evolve smoothly both before and after the moment of the maximum \citep[e.g.][]{nicholl21,gutierrez22}, which can be another difference between normal SNe-Ic and SLSNe-I. At first, \citet{lelo12} identified an early bump in the LC of SN~2006oz, which was followed by additional identifications of pre-peak undulations in the light, for example in the case of LSQ14bdq \citep{nicholl15b} and DES14X3taz \citep{smith16}. This discovery allowed \citet{nicholl16} to suggest that such bumps may be usual/omnipresent features in SLSNe-I, however, this theory was refuted by \citet{angus19}, who carried out a detailed analysis of SLSNe-I from the  Dark Energy Survey and found that only 3 SLSNe-I showed early bumps in their sample of 14 objects. Later on, PFT12dam \citep{vrees17}, and potentially SN~2018bsz \citep[e.g.][]{anderson18,chen21,pursi22,gutierrez22}  and SN~2018hti \citep{lin20, fiore22} were revealed to show early LC undulations. \citet{gutierrez22} examined this topic in detail, and found that some SLSNe-I that show smooth LC in their early phase, can be characterized by slow photometric evolution (e.g. SN~2007bi \citealt{galyam09,young10}, LSQ14an \citealt{inserra17}, SN~2015bn \citealt{nicholl16a,nicholl16} and SN~2017gci \citealt{fiore21} in their sample).

Numerous hypotheses have been put forward in the past decade to explain the possible physical causes of such pre-maximum LC undulations. According to \citet{lelo12}, the bumps may be caused by a recombination wave in the ejecta, while \citet{nicholl15} and \citet{piro15} suggests the post-shock cooling of the extended material that surrounds the progenitor. \citet{moriya12} built a model that assumes CSM-interaction, in which context, the pre-maximum bumps are caused by not excesses, but dips in the luminosity due to the increase of the optical depth in the CSM, when it becomes ionized. However, this scenario may be inconsistent with the velocities measured from the spectral features, since it requires a high covering fraction of the circumstellar material. \citet{kasen16} explains the bump by a magnetar-driven shock breakout, while \citet{margalit18} suggests the presence of a jet-driven wind that breaks through the ejecta. 

In can be seen in Table \ref{tab:osszefoglalo} that 6 objects in our sample show an early bump, and none of them belongs to the Type 15bn subgroup. Since the size of the sample is still limited, it cannot be concluded unambiguously that Type 15bn SLSNe-I never show bumps in their pre-maximum LCs, but there is a possibility that they occur in different environments, have different progenitors and explode in a different way compared to Type W SLSNe-I.

\subsection{Geometry}\label{subsec:geom}

It would be an interesting result to find that the difference between Type W and Type 15bn group can be found in the geometry of the explosion, therefore we examine our sample in this respect as well. The symmetry/asymmetry in the ejecta can be probed using polarimetry: if the observed SLSN shows null-polarization, the object is spherically symmetric, while polarization increasing with time implies that the inner part of the ejecta is more asymmetrical compared to the outer regions.

Unfortunately, polarimetric measurements were carried out only in case of 5 objects of our sample: 4 Type W (SN~2018hti, SN~2018bsz, LSQ14mo, and PTF12dam),  and 1 Type 15bn SLSN-I (SN~2015bn), as shown in Table \ref{tab:osszefoglalo}, therefore the examination of a larger sample is needed in the future to search for the connection between geometry and pre-maximum spectroscopic subtypes.

Here, we list the objects having polarimetric results in the literature.

\begin{itemize}
    \item {\bf SN~2018hti:} \citet{lee19} detected a polarization of 1.9 \% in case of this SLSN-I, however, it was revealed to originate from the interstellar dust around the object, suggesting that SN~2018hti shows null-polarization implying a symmetric explosion similarly to other Type I SLSNe examined until then.
    \item{\bf SN~2018bsz:} This SLSN-I was observed with polarimetry by \citet{maund21}, who found significant polarization of the object at $\sim$20 days rest-frame phase since maximum, which was explained to be due to a fundamental change in the asymmetry of the ejecta. According to \citet{inserra16,lelo17}, this phenomenon can be explained with an early-time emission that arises from an almost spherical outer layer, and then later the emission arises from a less symmetrical interior of the ejecta leading to the polarization increase with time. However, \citet{maund21} were able to obtain only single detection of polarization at $\sim$ 2 \%, thus it cannot be ruled out that the polarization was produced by the surrounding interstellar material.
    \item{\bf LSQ14mo:} \citet{lelo15b} examined this object with polarimetric techniques and found that the measured Stokes parameters did not evolve significantly in time. Although they showed a low value of polarization in the direction of LSQ14mo, it was interpreted as the polarization of the interstellar material. Thus LSQ14mo was concluded to be an explosion of a spherically symmetric ejecta.
    \item{\bf PTF12dam:}
    This SLSN-I was found to be a symmetric explosion showing null-polarization \citep{cikota18,pandey21,fred22}.
    \item{\bf SN~2015bn:} \citet{inserra16} presented the first spectropolarimetric measurements of SLSNe-I, nominally SN~2015bn, and found that this object shows significant polarization at +24 and +27 days phase relative to the maximum. After examining the Q - U plane, they found the presence of a main axis with no physical departure from that during the observed phases. They also revealed that the polarization increases with time, suggesting an axis-symmetric ellipsoidal configuration for the ejecta, which has a more asymmetric inner part compared to the outer regions. Consistently with this analysis, \citet{lelo17} confirmed the growth in asymmetry with imaging polarimetric observations using the Nordic Optical Telescope (NOT) at multiple epochs between -20 and +46 days phase.  They found that the interstellar polarization of the galaxy is negligible, and SN~2015bn shows an increase in polarization from $\sim$0.54\% to $>$ 1.10\% through the observed epochs. They dated the transition of the emission from the more symmetric outer layer to the asymmetric inner layer at $\sim$20 days after the maximum. In their two layers model, they suggested that the outer layer consists mostly of C and O, while the inner layer shows freshly synthesized, heavier elements.
  \item{\bf SN~2017egm:} Although this SLSN-I is not the member of the sample analyzed in this study, it is important to mention it here in regard of polarimetry, since it is a nearby object that had been analyzed by a lot of authors previously. In a recent study of \citet{fred22} it was also found to be a Type W SLSN-I. At first, \citet{bose18}  examined the host galaxy of SN~2017egm and found a $\sim$0.5 \% polarization that may be connected to global asymmetry. Later, \citet{maund19}  carried out polarimetric observations of SN~2017egm using the RINGO3 instrument of the Liverpool Telescope in 4 epochs. They concluded that the polarization measured in the direction of SN~2017egm is most probably caused by the interstellar material, not the SLSN itself. \citet{saito20}  carried out SUBARU observations in the later phases of the explosion of SN~2017egm and found that the angle of polarization differs significantly from the angle measured at earlier epochs. They corrected for the interstellar polarization and concluded that the objects were nearly symmetric in the early epoch (with a 0.2 \% polarization), which became asymmetric (0.8 \% polarization) later. 
  By the example of SN~2017egm, it is seen how complicated can be the polarimetry of SLSNe, but it is a very important field of topic that may reveal surprising results of the nature of SLSNe-I.
\end{itemize}

Since we know of only a few objects having polarimetric data, we cannot conclude that Type W  SLSNe-I  are "more symmetric" compared to Type 15bn  SLSNe-I . However, the continuation of examining SLSNe-I spectroscopically and polarimetrically may lead to the revelation of some connection between the geometry and the spectroscopic evolution of this kind of object. Therefore, polarimetric observations of SLSNe-I may open a new horizon for future studies.

\section{Summary}\label{sec:sum}

We have carried out spectroscopic analysis utilizing pre-maximum data of 27 Type I SLSNe-I obtained from the Open Supernova Catalog. Our main goal was to search for the reasons behind the pre-maximum spectroscopic differences of SLSNe-I, and find possible explanations for the presence of Type W and Type 15bn group disclosed by \citet{ktr21}. 

The members of Type W group were revealed to show a W-shaped feature in their pre-maximum spectra near 4000-5000 \AA, which is accommodated with the lines of O II. On the contrary, the objects of the other subtype, named after the representative of the group, SN~2015bn do now show such O II features in the early spectra. 

The SYN++ modeling of all available spectra revealed that Type W  SLSNe-I  tend to show larger photospheric temperature than 12000 K, while Type 15bn objects can be characterized by $T_{\rm phot} <$  12000 K, so Type W  SLSNe-I  are globally hotter compared to the other group. Consistently with previous studies, the C II and C III lines play a dominant role in the spectrum formation of both groups, suggesting the presence of a nearly pure C/O envelope without a significant amount of  higher elements originating from the inner layers of the ejecta. Oxygen features are present in the case of both groups as well, however, Type W SLSNe-I are generally show O II in contrast with Type 15bn SNe, which have only O I in their spectrum. The reason of this phenomenon may be the non-thermal excitation of O II, which requires a photospheric temperature larger than 12000 K \citep[e.g.][]{inserra19}. The borderline of 12000 K is supported by the other ions identified in the spectra as well: sodium, magnesium, and calcium features that need low photospheric temperature to be produced \citep{hatano99}, was identified only in Type 15bn  SLSNe-I , and were missing from the spectra of Type W SLSNe-I. Silicon and iron lines were found to be common in the case of both types, which is consistent with \citet{hatano99}, who showed that these elements may be present in the spectrum in a wide temperature range. 

 The two subtypes of SLSNe-I may be distinguished by the presence/absence of early bumps in the LC as well: 6 Type W SLSNe-I in our sample showed such LC undulations in the pre-maximum phase, while the LC of all Type 15bn SLSNe seemed to evolve smoothly before the maximum. However, as this sample contains only 9 Type 15bn SLSNe, the study of a larger sample is needed to show unambiguously that Type 15bn SLSNe do not show early bumps. 

Another difference of Type W and Type 15bn groups may be related to their geometry, however, only 5 objects in our sample had available polarimetric data. Our hypothesis is that Type W  SLSNe-I  may be spherically symmetric explosions, while Type 15bn  SLSNe-I  may show an increasing polarization with time, which suggests the presence of a spherically symmetric outer layer, and an asymmetric inner layer of the ejecta. This hypothesis is based on the fact that SN~2015bn shows increasing polarization in time, while objects Type W SN~2018hti, LSQ14mo, and PTF12dam were found to be symmetric previously. However, Type W SN~2018bsz may show an increase in polarimetry, based on a one-point measurement.

Therefore, further early phase polarimetric,  photometric and spectroscopic observations of SLSNe-I are needed to prove or disprove the connection between geometry and spectroscopic evolution. 

\acknowledgments
This study is supported by the \'UNKP-22-02 New National Excellence Program of the Ministry for Innovation and Technology. The author is thankful to the anonymous referee for the thorough report and useful comments that led to significant improvement of this paper.

{}

\section{Appendix}

\setcounter{table}{0}
\renewcommand{\thetable}{A\arabic{table}}

\setcounter{figure}{0}
\renewcommand{\thefigure}{A\arabic{figure}}

\begin{longtable}{lccccccccccccc}
\caption{Best-fit local parameter values of the  SYN++ modeling of the examined Type W  SLSNe-I  in the following units: $\log\tau$ (--),  $v_{\rm min}$ ($10^3$ km s$^{-1}$), $v_{\rm max}$ ($10^3$ km s$^{-1}$), aux ($10^3$ km s$^{-1}$), $T_{\rm exc}$ (1000 K).  } \\
\label{tab:local_w}
Ions & C II &  C III &  O I &  O II & O III & Na I & Mg II & Si II &  Si III & Ca II & Fe II &  Fe III &  Co III \\
\hline
  \multicolumn{14}{c}{SN~2018hti (-54 days phase)} \\
\hline

$\log\tau$ & -1.80 & -0.60 & -0.80 & -1.1 & & & & -1.60 & & & & & \\
$v_{\rm min}$ & 15.00 & 15.00 & 15.00 & 15.00 &&&& 20.00 &&&&&\\
$v_{\rm max}$ & 30.00 & 30.00 & 30.00 & 30.00 &&&& 30.00 &&&&&\\
aux & 3.00 & 2.00 & 2.00 & 2.00 &&&& 5.00 &&&&& \\
$T_{\rm exc}$ & 10.00 & 20.00 & 13.00 & 13.0 &&&& 3.00 &&&&& \\
\hline
 \multicolumn{14}{c}{DES14X3taz (-29 days phase)} \\
\hline
$\log\tau$ & & & &  -1.5 & -0.60 &&&&&& -0.50 && \\
$v_{\rm min}$ &&&& 13.00 & 13.00 &&&&&& 13.00 && \\
$v_{\rm max}$ &&&& 30.00 & 30.00 &&&&&& 30.00 && \\
aux &&&& 2.00 & 2.00 &&&&&& 1.00 &&\\
$T_{\rm exc}$ &&&& 13.0 & 15.00 &&&&&& 15.00 && \\
\hline
\multicolumn{14}{c}{SN~2010kd (-22 days phase)} \\
\hline
$\log\tau$ & -1.50 & -0.20 && -3.5 &&&& 0.70 &&&&&\\
$v_{\rm min}$ & 15.00 & 15.00 && 15.00 &&&& 22.00 &&&&& \\
$v_{\rm max}$ & 30.00 & 30.00 && 30.00 &&&& 30.00 &&&&&\\
aux & 3.00 & 2.00 && 2.00 &&&& 3.00 &&&&&\\
$T_{\rm exc}$ & 10.00 & 20.00 && 18.0 &&&& 12.00 &&&&& \\
\hline
\multicolumn{14}{c}{PTF09atu (-19 days phase)} \\
\hline
$\log\tau$ & -1.70 &&& -0.7 &&&& 1.20 &&& -2.30 && \\
$v_{\rm min}$ & 10.00 &&& 10.00 &&&& 20.00 &&& 10.00 && \\
$v_{\rm max}$ & 30.00 &&& 30.00 &&&& 30.00 &&& 30.00 &&\\
aux & 5.00 &&& 2.00 &&&& 2.00 &&& 2.00 &&\\
$T_{\rm exc}$ & 10.00 &&& 12.0 &&&& 12.00 &&& 12.00 && \\
\hline
\multicolumn{14}{c}{SN~2018bsz (-16 days phase)} \\
\hline
$\log\tau$ & -1.3 & 0.00 & -1.2 & -2.0 &&&&  &&& -0.40 && \\
$v_{\rm min}$ & 10.00 & 10.00 & 10.00 & 10.00 &&&&  &&& 10.00 && \\
$v_{\rm max}$ & 30.00 & 30.00 & 30.00 & 30.00 &&&&  &&& 30.00 &&\\
aux & 10.00 & 1.00 & 4.00 & 2.00 &&&&  &&& 1.00 &&\\
$T_{\rm exc}$ & 14.00 & 20.00 & 20.00 & 14.0 &&&&  &&& 12.00 &&\\
\hline
\multicolumn{14}{c}{PTF09cnd (-14 days phase)} \\
\hline
$\log\tau$ & -1.80 & -0.30 & -1.0 & -2.0 &&&&&&&&& \\
$v_{\rm min}$ & 13.00 & 13.00 & 13.00 & 13.00  &&&&&&&&&\\
$v_{\rm max}$ & 30.00 & 30.00 & 30.00 & 30.00  &&&&&&&&&\\
aux & 10.00 & 1.00 & 4.00 & 2.00  &&&&&&&&&\\
$T_{\rm exc}$ & 14.00 & 14.00 & 20.00 & 14.0  &&&&&&&&&\\
\hline
\multicolumn{14}{c}{LSQ14bdq (-11 days phase)} \\
\hline
$\log\tau$ &&&& -2.00 &&&&&&&&& \\
$v_{\rm min}$ &&&& 11.00 &&&&&&&&&\\
$v_{\rm max}$ &&&& 30.00 &&&&&&&&&\\
aux &&&& 2.00 &&&&&&&&&\\
$T_{\rm exc}$ &&&& 14.00&&&&&&&&& \\
\hline
\multicolumn{14}{c}{SN~2011kg (-10 days phase)} \\
\hline
$\log\tau$ & -1.70 & -0.70 && -0.4 &&&& -0.70 &&& -0.20 && \\
$v_{\rm min}$ & 11.50 & 11.50 && 11.50 &&&& 11.50 &&& 20.00 &&\\
$v_{\rm max}$ & 30.00 & 30.00 && 30.00 &&&& 30.00 &&& 30.00 &&\\
aux & 3.00 & 2.00 && 2.00 &&&& 15.00 &&& 2.00 &&\\
$T_{\rm exc}$ & 10.00 & 20.00 && 12.0 &&&& 12.00 &&& 22.00 &&\\
\hline
\multicolumn{14}{c}{SN~2006oz (-5 days phase)} \\
\hline
$\log\tau$ &&&& -1.7 &&&&&&& -0.70 && \\
$v_{\rm min}$ &&&& 15.00 &&&&&&& 15.00 && \\
$v_{\rm max}$ &&&& 30.00 &&&&&&& 30.00 &&\\
aux &&&& 2.00 &&&&&&& 1.00 &&\\
$T_{\rm exc}$ &&&& 14.0 &&&&&&& 15.00 &&\\
\hline
\multicolumn{14}{c}{iPTF13ajg (-5 days phase)} \\
\hline
$\log\tau$ && -0.20 && -2.2 &&&&&&& -0.40 && \\
$v_{\rm min}$ && 10.00 && 10.00 &&&&&&& 10.00 &&\\
$v_{\rm max}$ && 30.00 && 30.00 &&&&&&& 30.00 && \\
aux && 2.00 && 2.00 &&&&&&& 1.00 && \\
$T_{\rm exc}$ && 20.00 && 14.0 &&&&&&& 12.00 && \\
\hline
\multicolumn{14}{c}{SN~2016els (-5 days phase)} \\
\hline
$\log\tau$ & -1.00 & -0.20 & -0.8 & -1.6 &&&&& -0.3 &&&&\\
$v_{\rm min}$ & 13.00 & 13.00 & 13.00 & 13.00 &&&&& 20.00 &&&&\\
$v_{\rm max}$ & 30.00 & 30.00 & 30.00 & 30.00 &&&&& 30.00 &&&& \\
aux & 10.00 & 2.00 & 4.00 & 2.00 &&&&& 2.00 &&&&\\
$T_{\rm exc}$ & 20.00 & 20.00 & 20.00 & 14.0 &&&&& 18.00 &&&& \\
\hline
\multicolumn{14}{c}{SN~2019neq (-4 days phase)} \\
\hline
$\log\tau$ & -0.2 & -0.00 & 0.90 & -1.70 &&&&& -0.20 && -1.00 && -0.40 \\
$v_{\rm min}$ & 24.00 & 24.00 & 24.00 & 24.00 &&&&& 24.00 && 24.00 && 24.00 \\
$v_{\rm max}$ & 50.00 & 50.00 & 50.00 & 50.00 &&&&& 50.00 && 50.00 && 50.00 \\
aux & 1.00 & 2.00 & 1.00 & 1.00 &&&&& 2.00 && 2.00 && 2.00 \\
$T_{\rm exc}$ & 15.00 & 13.00 & 13.00 & 15.00 &&&&& 20.00 && 15.00 && 20.00 \\
\hline
\multicolumn{14}{c}{SN~2016ard (-4 days phase)} \\
\hline
$\log\tau$ & -1.00 & -0.20 & -0.8 & -1.6 &&&&&&& -1.70 && \\
$v_{\rm min}$ & 14.00 & 14.00 & 14.00 & 14.00 &&&&&&& 14.00 && \\
$v_{\rm max}$ & 30.00 & 30.00 & 30.00 & 30.00 &&&&&&& 30.00 &&\\
aux & 10.00 & 2.00 & 4.00 & 2.00 &&&&&&& 1.00 && \\
$T_{\rm exc}$ & 20.00 & 20.00 & 20.00 & 14.0 &&&&&&& 20.00 && \\
\hline
\multicolumn{14}{c}{SN~2005ap (-3 days phase)} \\
\hline
$\log\tau$ & -0.60 & -0.20 && -1.1 &&&&& -0.3 &&&& \\
$v_{\rm min}$ & 13.00 & 13.00 && 13.00 &&&&& 20.00 &&&& \\
$v_{\rm max}$ & 30.00 & 30.00 && 30.00 &&&&& 30.00 &&&& \\
aux & 10.00 & 2.00 && 2.00 &&&&& 2.00 &&&& \\
$T_{\rm exc}$ & 20.00 & 20.00 && 14.00 &&&&& 18.00 &&&&\\
\hline
\multicolumn{14}{c}{ Gaia16apd (-2 days phase)} \\
\hline
$\log\tau$ & -1.0 & -0.20 & -0.8 & -1.6 &&&&& 0.4 && -1.50 && \\
$v_{\rm min}$ & 20.00 & 20.00 & 20.00 & 20.00 &&&&& 20.00 && 20.00 && \\
$v_{\rm max}$ & 30.00 & 30.00 & 30.00 & 30.00 &&&&& 30.00 && 30.00 && \\
aux & 10.00 & 2.00 & 4.00 & 2.00 &&&&& 2.00 && 1.00 &&\\
$T_{\rm exc}$ & 20.00 & 20.00 & 20.00 & 14.0 &&&&& 18.00 && 20.00 &&\\
\hline
\multicolumn{14}{c}{SN~2010gx (-1 day phase)} \\
\hline
$\log\tau$ &&&& -1.50 &&&&& -0.0 &&&& \\
$v_{\rm min}$ &&&& 20.00 &&&&& 20.00 &&&&\\
$v_{\rm max}$ &&&& 30.00 &&&&& 30.00 &&&&\\
aux &&&& 2.00 &&&&& 2.00 &&&& \\
$T_{\rm exc}$ &&&& 14.00 &&&&& 18.00 &&&& \\
\hline
\multicolumn{14}{c}{LSQ14mo (-1 days phase)} \\
\hline
Ions & C II &  C III &  O I &  O II & O III & Na I & Mg II & Si II &  Si III & Ca II & Fe II &  Fe III &  Co III \\
\hline
$\log\tau$ && -0.80 && -2.1 &&&&&&& -0.60 && \\
$v_{\rm min}$ && 10.00 && 10.00 &&&&&&& 10.00  &&\\
$v_{\rm max}$ && 30.00 && 30.00 &&&&&&& 30.00 &&\\
aux && 2.00 && 2.00 &&&&&&& 1.00 &&\\
$T_{\rm exc}$ && 20.00 && 14.0 &&&&&&& 12.00 &&\\
\hline
\multicolumn{14}{c}{PTF12dam (-1 day phase)} \\
\hline
$\log\tau$ & -1.80 && -1.0 & -1.8 &&&&&&&&& \\
$v_{\rm min}$ & 11.00 && 11.00 & 11.00 &&&&&&&&&\\
$v_{\rm max}$ & 30.00 && 30.00 & 30.00 &&&&&&&&&\\
aux & 10.00 && 4.00 & 2.00 &&&&&&&&&\\
$T_{\rm exc}$ & 14.00 && 20.00 & 14.0 &&&&&&&&&\\
\hline
\end{longtable}

\begin{longtable}{lccccccccccccc}
\caption{Best-fit local parameter values of the  SYN++ modeling of the examined Type 15  SLSNe-I  in the following units: $\log\tau$ (--),  $v_{\rm min}$ ($10^3$ km s$^{-1}$), $v_{\rm max}$ ($10^3$ km s$^{-1}$), aux ($10^3$ km s$^{-1}$), $T_{\rm exc}$ (1000 K).  } \\
\label{tab:local_15bn}
Ions & C II &  C III &  O I &  O II & O III & Na I & Mg II & Si II &  Si III & Ca II & Fe II &  Fe III &  Co III  \\
\hline
  \multicolumn{14}{c}{PS1-14bn (-42 days phase)} \\
\hline
$\log\tau$ && -0.20 & 1.6 &&&&& -0.0 &&&& -0.5 & \\
$v_{\rm min}$ && 8.00 & 8.00 &&&&& 8.00 &&&& 8.00 &\\
$v_{\rm max}$ && 50.00 & 50.00 &&&&& 50.00 &&&& 50.00 &\\
aux && 6.00 & 1.00 &&&&& 6.00 &&&& 1.00 & \\
$T_{\rm exc}$ && 13.00 & 13.00 &&&&& 7.00 &&&& 7.00 &\\
\hline
\multicolumn{14}{c}{iPTF13ehe (-14 days phase)} \\
\hline
$\log\tau$ & -1.0 & -0.60 &&&&&& -0.2 &&&& -0.4 & \\
$v_{\rm min}$ & 10.00 & 10.00 &&&&&& 10.00 &&&& 10.00 &\\
$v_{\rm max}$ & 50.00 & 50.00 &&&&&& 50.00 &&&& 50.00 &\\
aux & 3.00 & 6.00 &&&&&& 6.00 &&&& 1 &\\
$T_{\rm exc}$ & 17.00 & 13.00 &&&&&& 7.00 &&&& 7.00 &\\
\hline
\multicolumn{14}{c}{SN~2018ibb (-11 days phase)} \\
\hline
$\log\tau$ & -1.4 && 0.30 &&&& 0.00 & 0.5 && -0.00 & 0.00 & -0.50 & \\
$v_{\rm min}$ & 8.00 && 8.00 &&&& 8.00, & 8.00 && 8.00 & 8.00 & 8.00  &\\
$v_{\rm max}$ & 50.00 && 50.00 &&&& 50.00 & 50.00 && 50.00 & 50.00 & 50.00 &\\
aux & 1.00 && 1.00 &&&& 1.00 & 1.00 && 1.00 & 1.00 & 1.00 &\\
$T_{\rm exc}$ & 10.00 && 10.00 &&&& 10.00 & 10.00 && 10.00 & 12.00 & 10.00 &\\
\hline
\multicolumn{14}{c}{PTF12gty (-4 days phase)} \\
\hline
$\log\tau$ & -0.2 & -0.1 & 0.60 &&&& 0.00 & 0.0 && 0.10 & 0.00  &&\\
$v_{\rm min}$ & 8.00 & 8.00 & 8.00 &&&& 8.00 & 8.00 && 8.00 & 8.00 && \\
$v_{\rm max}$ & 50.00 & 50.00 & 50.00 &&&& 50.00 & 50.00 && 50.00 & 50.00 && \\
aux & 1.00 & 2.00 & 2.00 &&&& 1.00 & 1.00 && 1.00 & 1.00 && \\
$T_{\rm exc}$ & 18.00 & 8.00 & 8.00 &&&& 10.00 & 10.00 && 10.00 & 12.00 && \\
\hline
\multicolumn{14}{c}{SSS120810 (-4 days phase)} \\
\hline
$\log\tau$ & -0.6 & -0.80 & 0.3 &&&&& -0.5 &&& -0.2 & -0.7 &\\
$v_{\rm min}$ & 10.00 & 10.00 & 10.00 &&&&& 10.00 &&& 10.00 & 10.00 &\\
$v_{\rm max}$ & 50.00 & 50.00 & 50.00 &&&&& 50.00 &&& 50.00 & 50.00 & \\
aux & 3.00 & 6.00 & 1 &&&&& 6.00 &&& 1.00 & 1.00 &\\
$T_{\rm exc}$ & 17.00 & 13.00 & 13.00 &&&&& 7.00 &&& 7.00 & 7.00 &\\
\hline
\multicolumn{14}{c}{SSS120810 (-4 days phase)} \\
\hline
$\log\tau$ & -0.8 & -0.60 & 0.4 & -0.7 &&&& -0.5 &&& 1 & -0.7 & \\
$v_{\rm min}$ & 10.00 & 10.00 & 10.00 & 10.00 &&&& 10.00 &&& 15.00 & 10.00 & \\
$v_{\rm max}$ & 50.00 & 50.00 & 50.00 & 50.00 &&&& 50.00 &&& 50.00 & 50.00 & \\
aux & 3.00 & 6.00 & 1 & 0.30 &&&& 6.00 &&& 1.00 & 1.00 &\\
$T_{\rm exc}$ & 17.00 & 13.00 & 13.00 & 13.00 &&&& 7.00 &&& 7.00 & 7.00 &\\
\hline
\multicolumn{14}{c}{ LSQ12dlf (-1 day phase)} \\
\hline
$\log\tau$ && -0.30 &&&&&& -0.7 &&& -0.0 & -0.1 & \\
$v_{\rm min}$ && 15.00 &&&&&& 15.00 &&& 15.00 & 15.00 & \\
$v_{\rm max}$ && 50.00 &&&&&& 50.00 &&& 50.00 & 50.00 & \\
aux && 6.00 &&&&&& 6.00 &&& 1.00 & 1.00 &\\
$T_{\rm exc}$ && 13.00 &&&&&& 7.00 &&& 7.00 & 7.00 &\\
\hline
\multicolumn{14}{c}{ LSQ12dlf (-1 day phase)} \\
\hline
$\log\tau$ & -1.0 & -0.3 & 0.50 &&&& 0.20 & 0.5 && 0.40 & 0.00 && \\
$v_{\rm min}$ & 7.00 & 7.00 & 7.00 &&&& 7.00 & 7.00 && 7.00 & 7.00 &&\\
$v_{\rm max}$ & 50.00 & 50.00 & 50.00 &&&& 50.00 & 50.00 && 50.00 & 50.00 &&\\
aux & 1.00 & 6.00 & 1.00 &&&& 1.00 & 1.00 && 1.00 & 1.00 &&\\
$T_{\rm exc}$ & 10.00 & 8.00 & 10.00 &&&& 10.00 & 8.00 && 10.00 & 12.00 && \\
\hline
\multicolumn{14}{c}{ LSQ14an (0 day phase)} \\
\hline
$\log\tau$ & -1.0 && 0.50 &&&& 0.00 & 0.5 && -0.00 & 0.30 & -0.50 & \\
$v_{\rm min}$ & 7.00  && 7.00 &&&& 7.00 & 7.00 && 7.00 & 7.00 & 7.00 &\\
$v_{\rm max}$ & 50.00  && 50.00 &&&& 50.00 & 50.00 && 50.00 & 50.00 & 50.00 &\\
aux & 1.00 && 6.00  &&&& 1.00 & 1.00 && 1.00 & 1.00 & 1.00 &\\
$T_{\rm exc}$ & 10.00  && 10.00 &&&& 10.00 & 8.00 && 10.00 & 12.00 & 10.00 &\\
\hline
\end{longtable}

\end{document}